\documentclass[12pt]{article}
\usepackage{amsfonts,amsmath,amsthm,amsxtra,amscd,amssymb}
\usepackage{txfonts,bm,pifont}
\usepackage{float}
\usepackage[dvipdfmx,hiresbb]{graphicx}
\usepackage[pdfencoding=auto,bookmarks=true,bookmarksnumbered=true,colorlinks=true]{hyperref}
\theoremstyle{plain}
\newtheorem{thm}{Theorem}[section]

\newtheorem{rem}[thm]{Remark}

\newcommand{\s}{\scriptstyle}
\newcommand{\mypicscale}{.42}
\newcommand{\mypicvin}{-0.3cm}
\newcommand{\mypichin}{-1.3cm}
\makeatletter
\@addtoreset{equation}{section} 

\makeatother
\begin{document}
\begin{center}
\begin{Large}
Integrable Discrete Model for One-dimensional Soil Water Infiltration\\[4mm]
\end{Large}
\begin{normalsize}
Dimetre {\sc Triadis} \\[2mm]
Institute of Mathematics for Industry, Kyushu University\\
744 Motooka, Fukuoka 819-0395, Japan\\
triadis@imi.kyushu-u.ac.jp\\[2mm]
Philip {\sc Broadbridge}\\[2mm]
Department of Mathematics and Statistics, La Trobe University\\
Bundoora, Victoria 3086, Australia\\
e-mail: P.Broadbridge@latrobe.edu.au\\[2mm]
Kenji {\sc Kajiwara}\\[2mm]
Institute of Mathematics for Industry, Kyushu University\\
744 Motooka, Fukuoka 819-0395, Japan\\
e-mail: kaji@imi.kyushu-u.ac.jp\\[2mm]
Ken-ichi {\sc Maruno}\\[2mm]
Department of Applied Mathematics, \\School of Fundamental Science and Engineering, 
Waseda University, \\
3-4-1 Okubo, Shinjuku-ku, Tokyo 169-8555, Japan\\
e-mail: kmaruno@waseda.jp\\[2mm]
\end{normalsize}
\end{center}
\begin{abstract}
We propose an integrable discrete model of one-dimensional soil water infiltration. This model is
based on the continuum model by Broadbridge and White, which takes the form of nonlinear
convection-diffusion equation with a nonlinear flux boundary condition at the surface. It is
transformed to the Burgers equation with a time-dependent flux term by the hodograph
transformation. We construct a discrete model preserving the underlying integrability, which is
formulated as the self-adaptive moving mesh scheme. The discretization is based on linearizability
of the Burgers equation to the linear diffusion equation, but the na\"ive discretization based on
the Euler scheme which is often used in the theory of discrete integrable systems does not
necessarily give a good numerical scheme. Taking desirable properties of a numerical scheme into
account, we propose an alternative discrete model that produces solutions with similar accuracy to
direct computation on the original nonlinear equation, but with clear benefits regarding
computational cost.
\end{abstract}

\section{Introduction}
With the volumetric water content $\theta$ adopted as the dependent variable, the Richards equation
for flow of water through unsaturated soil is given in the form of a nonlinear diffusion-convection
equation (e.g. \cite{Smith,Warrick})
\begin{equation}
\theta_t=\nabla\cdot\left[D(\theta)\nabla\theta\right]-K'(\theta)\theta_z,
\label{Richards}
\end{equation}
where $t$ represents time, $z$ is the depth coordinate, $K$ is the hydraulic conductivity and $D$ is
the soil-water diffusivity.  Over the past 60 years, there have been developed many analytic and
numerical schemes to construct exact and approximate solutions to (\ref{Richards}), subject to
meaningful boundary conditions on geometric domains of practical interest at the laboratory, field
or regional scales \cite{Smith,Warrick}.  There are a number of useful integrable models
$\left(K(\theta),D(\theta)\right)$ for unsteady flows in one dimension or steady flows in higher
dimensions. The current study will develop associated integrable finite difference models on a
space-time grid.

Discretization of soliton equations preserving integrability has been studied actively, after the
pioneering work of Ablowitz--Ladik
\cite{Ablowitz-Ladik:dNLS_1976,Ablowitz-Ladik:dNLS_1977,Ablowitz:book} and Hirota
\cite{Hirota:difference1,Hirota:difference2,Hirota:dsG,Hirota:difference4,Hirota:difference5}.  Some
time afterwards, Date, Jimbo and Miwa developed a unified algebraic approach from the view of
so-called the KP theory
\cite{DJM:discrete1,DJM:discrete2,DJM:discrete3,DJM:discrete4,DJM:discrete5,Jimbo-Miwa,Miwa}.  In
recent decades discrete integrable systems have been used as a theoretical background or testbed for
constructing good discrete models. For example, they have been used as a foundation for the study of
discrete curves and surfaces known as discrete differential geometry, which has wide application,
for example in computer graphics \cite{book:Bobenko_Suris}.  Nishinari--Takahashi considered the
Burgers equation as a traffic model and constructed discrete and ultradiscrete integrable models,
through which they gave a unified view to various continuous, discrete and cellular automaton
traffic models \cite{Nishinari_Takahashi:Burgers}. For further recent developments in discrete
integrable systems, see for example \cite{book:HJN,KNY:Painleve_Review,book:Suris}.

It should be noted that most studies of discrete integrable systems have been theoretical because of
their underlying rich mathematical structures, but originally they were studied from a need for
stable and accurate numerical computations for soliton equations, with the expectation that
underlying integrability, in particular a sufficient number of conserved quantities, would
contribute to numerical stability and accuracy \cite{HA2,ASH2,ASH1,TA1,HA1,TA2,TA3,TA4}. However,
there are not so many examples where discrete integrable models have been used to simulate real
problems.

In this paper, we consider an integrable model for soil water infiltration, formulated as a
nonlinear diffusion-convection equation with a nonlinear flux boundary condition.  This equation is
reducible to a nonlinear boundary value problem of the Burgers equation with a boundary flux that
results from the hodograph transformation, an independent variable transformation including the
dependent variable. Furthermore, the Burgers equation is reduced to the linear diffusion equation by
the Cole-Hopf transformation. We then construct a discrete model with these properties being
preserved. Amazingly, the resulting numerical scheme is formulated as a self-adaptive moving mesh
scheme which has been proposed in the study of numerical schemes for nonlinear wave equations (for
example, the Camassa-Holm equation and the short pulse equation) related to hodograph
transformations \cite{FMO:CH,FMO:SP1,FMO:SP2}. Practical variable-flux boundary conditions may be
readily and naturally adopted in the proposed discrete model; even in the integrable continuum
model, general time-dependent flux boundary conditions lead to unresolved mathematical difficulties.

Discretization of integrable systems relies on the underlying linear structure. In the case of the
Burgers equation, discretization is carried out so that linearizability to the diffusion equation is
preserved \cite{HerederoLeviWinternitz1999,Hirota:difference5,Nishinari_Takahashi:Burgers}. However,
the actual discretization of the linear equation is usually chosen without paying attention to
properties of a numerical scheme.  From a viewpoint separated from integrability, we show that we
must consider the numerical stability of discretizations to produce applicable discrete models.

This paper is organized as follows. In Section \ref{sec:continuous_model} we give an integrable
model of one-dimensional soil water infiltration \cite{BroadbridgeWhite1988a} and its
transformations to the Burgers and linear diffusion equations. In Section \ref{sec:discrete} we
construct discrete models preserving integrability; a model based on the standard Euler scheme for
linear diffusion equation in Section \ref{subsec:discrete_Euler}, and an alternative model based on
the Crank--Nicolson scheme in Section \ref{subsec:discrete_Crank-Nicolson}.  We show that the former
model has built-in numerical instability, while the latter model provides us with a stable and
reasonably accurate numerical scheme. Section \ref{subsec:direct_comparison} compares the
performance of the Crank-Nicolson integrable model with the Crank-Nicolson scheme applied directly
to our original nonlinear diffusion-convection equation. Concluding remarks are given in Section
\ref{sec:Concluding_Remarks}.
\section{An integrable model for soil water infiltration}\label{sec:continuous_model}
We consider the following initial-boundary value problem of a one-dimensional convection-diffusion
equation for $\theta=\theta(z,t)$ \cite{BroadbridgeWhite1988a}
\begin{align} 
&\frac{\partial\theta}{\partial t}=-\frac{\partial}{\partial z}
  \left[\frac{\lambda}{2(b-\theta)}+\gamma(b-\theta)+\beta-\frac{a}{(b-\theta)^2}\frac{\partial\theta}{\partial z}\right],
\label{eqn:continuous_model_theta0}
\\
& z\ge0,\quad t\ge0,
\nonumber\\[2mm]
& \theta(z,0) = \theta^{(0)}(z), 
\nonumber\\
& J(0,t)= \left.\frac{\lambda}{2(b-\theta)}+\gamma(b-\theta) + \beta
- \frac{a}{(b-\theta)^2}\frac{\partial\theta}{\partial z} \right|_{z=0} = R(t),
\label{eqn:continuous_model_theta_ic_bc0}
\\ 
& \lim_{z\to\infty}\theta(z,t) = \lim_{z\to\infty}\theta^{(0)}(z)=\theta^{(n)}.
\nonumber
\end{align}
Here, $\theta(z,t)$ is volumetric water content of soil, $\theta^{(0)}(z)$ is a given function, for
the present study $\theta^{(n)}$ is taken to be \hbox{$\min \limits_{z}\theta^{(0)}(z) =
\lim\limits_{z\to\infty}\theta^{(0)}(z)$}, $J(z,t)$ is water flux density, and $a$, $b$, $\beta$,
$\gamma$, $\lambda$ are parameters.  This is a special case of the Richards equation
(\ref{Richards}) with
\begin{equation}
\nabla = \frac{\partial}{\partial z},\quad  
D(\theta) = \frac{a}{(b-\theta)^2}, \quad 
K(\theta) = \frac{\lambda}{2(b-\theta)} + \gamma(b-\theta) +\beta,
\end{equation}
which describes one-dimensional soil water infiltration with specified water flux $R(t)$ at
the surface $z=0$. These special functional forms of the diffusivity $D(\theta)$ and hydraulic
conductivity $K(\theta)$ ensure that the Richards equation is linearisable, but are general enough
to model a range of real soils \cite{WhiteBroadbridge1988}.

It is possible to normalize $\theta$ as $0\leq \theta(z,t)\leq 1$ by replacing $\theta(z,t)$ by
\hbox{$[\theta(z,t)-\theta^{(n)}]/[\theta^{(s)}-\theta^{(n)}]$}, where $\theta^{(s)}$ is the
saturated volumetric water content. Further, applying suitable scale changes, we can adopt the
dimensionless variables and parameters normalized as in \cite{BroadbridgeWhite1988a}:
\begin{equation}
\begin{array}{lll}\smallskip
{\displaystyle a = C(C-1),} & {\displaystyle b=C,} &\\
{\displaystyle \lambda = 2 C^2(C-1),}  &{\displaystyle \gamma = C-1,} &{\displaystyle \beta = -2 C( C-1).}
\end{array}\label{eqn:continuous_model_parametrization}
\end{equation}
Here $C>1$ is a characteristic parameter of the soil describing the strength of concentration-dependence of
hydraulic properties, typically 1.02 (strong) $\sim$ 1.5 (weak).
The model is parametrized by the
single parameter $C$, but we use $a$, $b$ and $\beta$ for notational simplicity. Then we consider
the normalized model 
\begin{equation} 
 \frac{\partial\theta}{\partial t}=-\frac{\partial}{\partial z}
  \left[\frac{ab}{b-\theta} + \frac{a}{b}(b-\theta)+\beta-\frac{a}{(b-\theta)^2}\frac{\partial\theta}{\partial z}\right],
\label{eqn:continuous_model_theta}
\end{equation}
\begin{equation}
\begin{split}
& z\ge0,\quad t\ge0,\\[2mm]
& \theta(z,0) = \theta^{(0)}(z), \\
& J(0,t)= \left.\frac{ab}{b-\theta} + \frac{a}{b}(b-\theta) + \beta
- \frac{a}{(b-\theta)^2}\frac{\partial\theta}{\partial z} \right|_{z=0} = R(t),\\[4mm]
& \lim_{z\to\infty}\theta(z,t) =\lim_{z\to\infty}\theta^{(0)}(z) =  0.
\end{split}
\label{eqn:continuous_model_theta_ic_bc}
\end{equation}

The model \eqref{eqn:continuous_model_theta}, \eqref{eqn:continuous_model_theta_ic_bc} is {\em
integrable} in a sense that it is transformed to the celebrated {\em Burgers equation} and thus
linearizable by suitable change of variables.  To demonstrate this, we first apply the dependent
variable transformation called the {\em Kirchhoff transformation} \cite{Kirchhoff}
\begin{equation}
 \mu=\int D(\theta)\,d\theta = \frac{a}{b-\theta}, \label{eqn:Kirchhoff_transformation}
\end{equation}
after which \eqref{eqn:continuous_model_theta} is written as
\begin{equation}
 \frac{\partial \mu}{\partial t}= \frac{\mu^2}{a}\frac{\partial^2 \mu}{\partial z^2} 
+ \left[ \frac{a}{b} -  \frac{b\mu^2 }{a}\right] \frac{\partial \mu}{\partial z}.
\label{eqn:continuous_model_mu_zt}
\end{equation}
We next apply the independent variable transformation called the {\em Storm transformation} \cite{Storm1951}
\hbox{$(z,t)\rightarrow (Z,\tau)$}
\begin{equation}
Z = a^{\frac{1}{2}}\int_0^z\frac{1}{\mu(z,t)}\,dz \;
\left( \; \text{or}\; z = a^{{-1}/{2}} \int_0^Z \mu(Z,\tau)\, dZ \; \right) \; ; \; 
\tau=t.
\label{eqn:Storm_transformation}
\end{equation}
This transforms \eqref{eqn:continuous_model_mu_zt} to
\begin{equation}
\frac{\partial\mu}{\partial\tau}=\frac{\partial^2\mu}{\partial Z^2}-2ba^{-1/2}\mu\frac{\partial\mu}{\partial Z}+a^{-1/2}\Big[R(\tau)-\beta\Big]\frac{\partial\mu}{\partial Z}.
\label{eqn:continuous_model_mu_Ztau}
\end{equation}
The initial and boundary conditions \eqref{eqn:continuous_model_theta_ic_bc} are transformed to
\begin{equation}
\begin{array}{c}\medskip
{\displaystyle  \mu(Z,0) = \mu^{(0)}(Z)=\frac{a}{b-\theta^{(0)}(Z)},}\\
\bigskip
{\displaystyle J(0,\tau) = \left. b\mu + \frac{a^2}{b\mu}+\beta - a^{\frac{1}{2}}\frac{\mu_Z}{\mu}\right|_{Z=0} = R(\tau),}\\
{\displaystyle \lim_{Z\to\infty} \mu(Z,\tau) = \frac{a}{b},}
\end{array}
\label{eqn:continuous_model_mu_ic_bc}
\end{equation}
respectively. Equation \eqref{eqn:continuous_model_mu_Ztau} is essentially the Burgers equation,
where the third term in the right-hand side originates from the surface boundary condition. We
remark that the Storm transformation \eqref{eqn:Storm_transformation} is nothing but the hodograph
(reciprocal) transformation \cite{CourantFriedrichs1948} associated with the conserved density $1/\mu$ of
\eqref{eqn:continuous_model_mu_zt}, or $\mu$ of \eqref{eqn:continuous_model_mu_Ztau}.  Note that the
boundary condition as $z \to \infty$ corresponds to the condition as $Z \to \infty$ due to
\eqref{eqn:Kirchhoff_transformation} and \eqref{eqn:Storm_transformation}, since $1/\mu(z,t)$ does
not become asymptotically $0$ as $z\to\infty$ in general. Practically we may impose this condition
at sufficiently large $Z$.

It is well-known that the Burgers equation admits linearization by the {\em Cole--Hopf transformation}
\begin{equation}
 \mu = -\frac{a^{\frac{1}{2}}}{b}\,\frac{1}{\phi}\frac{\partial \phi}{\partial Z}. \label{eqn:Cole-Hopf_transformation}
\end{equation}
Then \eqref{eqn:continuous_model_mu_zt} is reduced to the linear diffusion equation
\begin{equation}
\frac{\partial \phi}{\partial \tau} = \frac{\partial^2\phi}{\partial Z^2}
+ \frac{1}{a^{\frac{1}{2}}}\, \Big[R(\tau)-\beta\Big]\,\frac{\partial \phi}{\partial Z}.\label{eqn:continuous_model_phi_Ztau}
\end{equation}
Let us write down the initial and boundary conditions for $\phi$. The initial condition in
\eqref{eqn:continuous_model_mu_ic_bc} and \eqref{eqn:Cole-Hopf_transformation} gives
\begin{equation}
\mu^{(0)}(Z) = \left.-\frac{a^{\frac{1}{2}}}{b}\,\frac{1}{\phi}\frac{\partial \phi}{\partial Z}\right|_{\tau=0}, 
\label{eqn:continuous_model_phi_ic_eq}
\end{equation}
which is integrated as
\begin{equation}
  \phi(Z,0) = \exp\left[-\frac{b}{a^{\frac{1}{2}}}\int_0^Z \mu^{(0)}(Z)\, dZ\right]. \label{eqn:continuous_model_phi_ic_sol}
\end{equation}
The flux $J(Z,\tau)$ is rewritten in terms of $\phi$ by using \eqref{eqn:Cole-Hopf_transformation}
as
\begin{equation}
\begin{split}
 J(Z,\tau) &= b\mu + \frac{a^2}{b\mu}+\beta - a^{\frac{1}{2}}\frac{\mu_Z}{\mu}
 = -\frac{a^{\frac{1}{2}}}{\phi_Z}\left(\phi_{ZZ} + a\phi - \frac{\beta}{a^{\frac{1}{2}}}\phi_Z\right)\\
&=-\frac{a^{\frac{1}{2}}}{\phi_Z}\left(\phi_{\tau} + a\phi - \frac{R(\tau)}{a^{\frac{1}{2}}} \phi_Z \right), 
\end{split}\label{eqn:continuous_flux_phi}
\end{equation}
where we have used the differential equation \eqref{eqn:continuous_model_phi_Ztau}.  Then the
boundary condition at $Z=0$ in \eqref{eqn:continuous_model_mu_ic_bc} gives
\begin{equation}
\left.\frac{\partial\phi}{\partial \tau} + a\phi\right|_{Z=0}=0, \label{eqn:continuous_model_phi_bc_eq0}
\end{equation}
which is integrated as
\begin{equation}
 \phi(0,\tau) = e^{-a\tau}.\label{eqn:continuous_model_phi_bc_sol0} 
\end{equation}
The boundary condition of $\phi$ for large $Z$ in \eqref{eqn:continuous_model_mu_ic_bc} yields
by using \eqref{eqn:Cole-Hopf_transformation}
\begin{equation}
\frac{a}{b}=-\frac{a^{\frac{1}{2}}}{b}\,\frac{1}{\phi}\frac{\partial \phi}{\partial Z},
\end{equation}
which is integrated as
\begin{equation}
 \phi(Z,\tau) = g(\tau)\,e^{-a^{1/2} Z},\; {\rm as} \; 
Z\to\infty,\label{eqn:continuous_model_phi_bc_infty0}
\end{equation}
where $g(\tau)$ is an arbitrary function to be determined from consistency with the initial condition. 
Substituting \eqref{eqn:continuous_model_phi_bc_infty0} into \eqref{eqn:continuous_model_phi_Ztau}, we find that
$g(\tau)$ satisfies
\begin{equation}
 g_\tau = -\left(a + R(\tau) \right)g,
\end{equation}
so that 
\begin{equation}
 g(\tau) = g_0\exp\left[- a\tau - \int_0^\tau R(s)\,ds\right],
\end{equation}
and
\begin{equation}
 \phi(Z,\tau) = g_0\exp\left[-a^{\frac{1}{2}}Z - a\tau - \int_0^\tau R(s)\,ds\right], \; {\rm as} \; Z \to \infty,
\end{equation}
where $g_0$ is a constant to be determined from consistency with the
initial condition \eqref{eqn:continuous_model_phi_ic_sol}. Finally we have
\begin{equation}
\phi(Z,\tau) = 
\exp\left[ -\frac{b}{a^{\frac{1}{2}}}\int_0^Z \mu^{(0)}(Z)\, dZ
- a\tau - \int_0^\tau R(s)\,ds \right], \; {\rm as}\; Z\to\infty.
\label{eqn:continuous_model_phi_bc_infty}
\end{equation}

Summarizing the discussion above, we obtain the following ``equivalent'' three models:\\[2mm]
\noindent (i) Original model: \eqref{eqn:continuous_model_parametrization},
\eqref{eqn:continuous_model_theta}, \eqref{eqn:continuous_model_theta_ic_bc}.\\
\noindent (ii) Burgers model: \eqref{eqn:continuous_model_mu_Ztau},
\eqref{eqn:continuous_model_mu_ic_bc}.\\
\noindent (iii) Linear model: \eqref{eqn:continuous_model_phi_Ztau},
\eqref{eqn:continuous_model_phi_ic_sol}, \eqref{eqn:continuous_model_phi_bc_sol0},
\eqref{eqn:continuous_model_phi_bc_infty}.\\[2mm]
Note that (i) and (ii) are related by \eqref{eqn:Kirchhoff_transformation} and
\eqref{eqn:Storm_transformation}, (ii) and (iii) by \eqref{eqn:Cole-Hopf_transformation}.

It may be useful to write down the initial and boundary conditions specialized to an initial
condition of practical importance
\begin{equation}
 \theta^{(0)}(z) = 0,\quad \mu^{(0)}(Z)=\frac{a}{b}.\label{eqn:continuous_zero_ic}
\end{equation}
Then \eqref{eqn:continuous_model_phi_ic_sol}, \eqref{eqn:continuous_model_phi_bc_sol0} and 
\eqref{eqn:continuous_model_phi_bc_infty} become
\begin{equation}
\begin{split}
& \phi(Z,0) = e^{-a^{1/2}Z} ,\quad
 \phi(0,\tau) = e^{-a\tau}, \\
&\phi(Z,\tau) = 
\exp\left[ -a^{\frac{1}{2}}Z - a\tau - \int_0^\tau R(s)\,ds \right], \; {\rm as}\; Z\to\infty,
\end{split}
\end{equation}
respectively.

\section{Integrable discrete models}\label{sec:discrete}
In this section, we consider a full discretization (discretization in both space and time) of the
model discussed in Section \ref{sec:continuous_model}. Integrable discretization of soliton
equations has been actively studied for a long time \cite{book:grammati,book:HJN,book:Suris}.  In
particular, the discretization of the Burgers equation has been carried out preserving
linerizability in \cite{Hirota:difference5}, and used to model traffic in
\cite{Nishinari_Takahashi:Burgers} after application of so-called {\em ultradiscretization} to
construct a cellular automaton model. In \cite{HerederoLeviWinternitz1999} symmetry of the discrete
Burgers equation is discussed.
\subsection{Discrete Burgers and linear models}\label{subsec:discrete_Euler}
We start with discretization of the linear model
\eqref{eqn:continuous_model_phi_Ztau}, \eqref{eqn:continuous_model_phi_ic_sol},
\eqref{eqn:continuous_model_phi_bc_sol0} and \eqref{eqn:continuous_model_phi_bc_infty}.  Putting 
\begin{align}
 \phi(Z,\tau)&=\phi(n\epsilon,m\delta)=\phi_n^m,\quad
R(\tau)=R(m\delta)=R^m,
\nonumber \\
& \quad n=1,2,\ldots,N,\quad m=0,1,2\ldots,
\end{align}
with $\epsilon$, $\delta$ being lattice intervals of $n$ and $m$, respectively, let
us consider the following partial difference equation as a discretization of
\eqref{eqn:continuous_model_phi_Ztau}:
\begin{equation}
\begin{split}
& \frac{\phi^{m+1}_{n} - \phi^m_{n}}{\delta}
=  \frac{\phi^m_{n+1} - 2 \phi^m_{n} + \phi^m_{n-1}}{\epsilon^2} 
+  \frac {R^m - \beta}{a^{\frac{1}{2}} }\, \frac{\phi^m_{n+1} - \phi^m_{n-1} }{2 \epsilon},\\[2mm]
&\hspace*{40pt} n=2,\ldots,N-1, \quad m=0,1,2,\ldots\,.
\end{split}
\label{eqn:dBurgers1}
\end{equation}
We note that $R^m$ plays the role of the given discrete surface flux as in the continuous model. We
next consider discretization of the Cole--Hopf transformation \eqref{eqn:Cole-Hopf_transformation}.
Here we adopt
\begin{equation}
 \mu_n^m = -\frac{2a^{\frac{1}{2}}}{b\epsilon}\, 
\frac{\phi_{n+1}^m-\phi_n^m}{\phi_{n+1}^m + \phi_n^m}.
\label{eqn:discrete_Cole-Hopf_transformation}
\end{equation}
\begin{rem}\label{etaremark} 
The choice of \eqref{eqn:discrete_Cole-Hopf_transformation} may be justified as follows. Consider
Taylor series expansions of $ \phi_{n+1}^m$ and $\phi_{n}^m$ about the point $\phi\big((n+\frac
12)\epsilon, m \delta\big) = \phi^m_{n+1/2}$. We find that
\begin{equation}
\mu^m_n =  -\frac{a^{\frac{1}{2}}}{b} 
\left. \frac
{\textstyle \frac { \partial \phi}{\partial Z} }
{\phi}
\right|_{Z=(n+\frac 12)\epsilon} + O\big(\epsilon^2\big),
\end{equation}
so that the Cole-Hopf transformation is of second order in space if we associate position $Z = (n+\frac 12)\epsilon$ with $\mu^m_n$.
\end{rem}

We proceed to discretization of the initial condition. By using \eqref{eqn:discrete_Cole-Hopf_transformation}, 
equation \eqref{eqn:continuous_model_phi_ic_eq} may be discretized as
\begin{equation}
\mu^{(0)}_n = -\frac{2a^{\frac{1}{2}}}{b\epsilon}\, \frac{\phi_{n+1}^0-\phi_n^0}{\phi_{n+1}^0 + \phi_n^0}.
\label{eqn:discrete_model_phi_ic_eq}
\end{equation}
Here, $\mu^{(0)}_n$ is a given function in $n$ which will play the role of the initial value of the
discrete counterpart of the Burgers model. Equation \eqref{eqn:discrete_model_phi_ic_eq} 
can be explicitly solved as
\begin{equation}
%
%
\phi_n^0 = \prod_{j=0}^{n-1}P_j,\quad P_j = 
\frac{1-\frac{b\epsilon}{2a^{1/2}}\,\mu_j^{(0)}}{1+\frac{b\epsilon}{2a^{1/2}}\,\mu^{(0)}_j}.
 \label{eqn:discrete_model_phi_ic_sol} 
\end{equation}

We next consider the boundary conditions. We can impose the surface boundary condition at $n=1$ by a
simple discretization of \eqref{eqn:continuous_model_phi_bc_eq0} \footnote{Imposing the boundary
condition at $n=1$ but not at $n=0$ is due to a technical reason to avoid introducing a virtual
value $\phi_{-1}^m$.}:
\begin{equation}
 \frac{\phi_1^{m+1} - \phi_1^{m}}{\delta} = -a\phi_1^m,
\label{eqn:discrete_model_phi_bc_eq1}
\end{equation}
which is integrated as
\begin{equation}
 \phi_1^m = \phi^{(0)}\,\left(1 -a\delta\right)^m.\label{eqn:discrete_model_phi_bc_sol1}
\end{equation}
Here, $\phi^{(0)}$ is a constant to be determined from the consistency with the initial condition.
Actually, putting $m=0$ in \eqref{eqn:discrete_model_phi_bc_sol1} and comparing with
\eqref{eqn:discrete_model_phi_ic_sol}, we have
\begin{equation}
 \phi_1^m  =  
P_0\,\left(1 - a\delta\right)^m. \label{eqn:discrete_model_phi_bc_sol}
\end{equation}
Comparing with the continuous case, the boundary condition at $n=N$ consistent with the initial
condition may be written in the form
\begin{equation}
 \phi_N^m = g^m\prod_{j=0}^{N-1}P_j, \label{eqn:discrete_model_phi_bc_infty0}
\end{equation}
where $g^m$ is a function of $m$ to be determined as follows: substituting
\eqref{eqn:discrete_model_phi_bc_infty0} into \eqref{eqn:dBurgers1} with $n=N-1$ we have
\begin{equation}
g^{m+1} = \left[1+
\frac{\delta}{\epsilon^2}\left(
P_{N} - 2 + \frac{1}{P_{N-1}}
\right)
+ \frac{\delta}{\epsilon}\frac{R^m-\beta}{a^{\frac{1}{2}}}
\left(P_{N} - \frac{1}{P_{N-1}}\right)\right]g^m,
\end{equation}
so that
\begin{equation}
  \phi_N^m = \prod_{j=0}^{N-1}P_j \prod_{i=0}^{m-1}\left[1+
\frac{\delta}{\epsilon^2}\left(
P_{N} - 2 + \frac{1}{P_{N-1}}
\right)
+ \frac{\delta}{\epsilon}\frac{R^i-\beta}{a^{\frac{1}{2}}}
\left(P_{N} - \frac{1}{P_{N-1}}\right)\right].
\label{eqn:discrete_model_phi_bc_infty}
\end{equation}
Therefore, the discrete linear model is formulated as \eqref{eqn:dBurgers1} with initial condition
\eqref{eqn:discrete_model_phi_ic_sol} and boundary conditions \eqref{eqn:discrete_model_phi_bc_sol},
\eqref{eqn:discrete_model_phi_bc_infty}. 
\begin{rem}
In practical numerical computation, the boundary condition at $n=N$
\eqref{eqn:discrete_model_phi_bc_infty} is incorporated simply as follows.  At fixed $m$, the boundary value $\phi_1^m$ is given by \eqref{eqn:discrete_model_phi_bc_sol}, and $\phi_n^m$ for $n=2,3,\ldots N-1$
are computed successively by \eqref{eqn:dBurgers1} using $\phi_n^{m-1}$ ($n=1,\ldots,N$).  Then
$\phi_N^m$ is determined by $\phi_N^m = P_{N-1}\phi_{N-1}^m$, instead of evaluating
\eqref{eqn:discrete_model_phi_bc_infty} directly, under the assumption that the simulation time is
not large enough for the large-z initial condition to be perturbed.
\end{rem}
Now that we have $\phi_n^m$ ($n=1,\ldots,N$, $m=0,1,\ldots$), $\mu_n^m$ and $\theta_n^m$ are given
by \eqref{eqn:discrete_Cole-Hopf_transformation} and
\begin{equation}
\theta_n^m = b - \frac{a}{\mu_n^m}
= b\left(1  + \frac{a^{\frac{1}{2}}\epsilon}{2}\,
\frac{\phi_{n+1}^m + \phi_n^m}{\phi_{n+1}^m-\phi_n^m}\right),\label{eqn:discrete_model_theta}
\end{equation}
respectively, for $n=1,\ldots N-1$. $\mu_0^m$ and $\theta_0^m$ are obtained as follows. Consider the
linear equation \eqref{eqn:dBurgers1} at $n=1$
\begin{equation}
\frac{\phi^{m+1}_{1} - \phi^m_{1}}{\delta}
=  \frac{\phi^m_{2} - 2 \phi^m_{1} + \phi^m_{0}}{\epsilon^2} 
+  \frac {R^m - \beta}{a^{\frac{1}{2}} }\, \frac{\phi^m_{2} - \phi^m_{0} }{2 \epsilon}.
\label{eqn:dBurgers1_at_n=1}
\end{equation}
Here, $\phi_1^m$, $m=0,1,2,\ldots$ are given in \eqref{eqn:discrete_model_phi_bc_sol}.  Dividing the
both side of \eqref{eqn:dBurgers1_at_n=1} by $\phi_1^m$ and introducing the auxiliary dependent
variable $u_n^m$ by
\begin{equation}
 u_n^m = \frac{\phi_{n+1}^m}{\phi_n^m},\label{eqn:u}
\end{equation}
we find that unknown variable $u_0^m$ can be computed from known $u_1^m$ as
\begin{equation}
 u_0^m = \frac{1-\kappa^m}{2 - a\epsilon^2 - (1+\kappa^m)u_1^m},\quad \kappa^m 
= \frac{\epsilon(R^m-\beta)}{2a^{\frac{1}{2}}}. \label{eqn:u0}
\end{equation}
Here, we used \eqref{eqn:discrete_model_phi_bc_sol} so that $\phi_1^{m+1}/\phi_1^m = 1-a\delta$.
Then $\mu_0^m$ and $\theta_0^m$ are computed as
\begin{equation}
 \mu_0^m =  -\frac{2a^{\frac{1}{2}}}{b\epsilon}\, 
\frac{u_{0}^m-1}{u_{0}^m + 1},\quad
\theta_0^m = b\left(1 + \frac{a^{\frac{1}{2}}\epsilon}{2}\, 
\frac{u_{0}^m + 1}{u_{0}^m-1}\right).\label{eqn:mu0_by_u0}
\end{equation}
Hence we obtain $\mu_n^m$ and $\theta_n^m$ for $n=0,\ldots,N-1$, $m=0,1,2,\ldots$.

$\theta_n^m$ in \eqref{eqn:discrete_model_theta} corresponds to $\theta(Z,\tau)$ in the continuous
model.  In order to obtain $\theta(z,t)$, we have to construct and apply the discrete version of
hodograph transformation \eqref{eqn:Storm_transformation}. Discretization of the hodograph
transformation has already appeared in the study of numerical schemes (which are called
self-adaptive moving mesh schemes) for nonlinear wave equations such as the Camassa-Holm equation
and the short pulse equation \cite{FMO:CH,FMO:SP1,FMO:SP2} and the dynamics of discrete planar
curves \cite{hodograph,FIKMO}, and as a consequence, one may simply replace the integration in
\eqref{eqn:Storm_transformation} by summation. Practically, we may use the trapezoidal rule so that
the precision is $O(\epsilon^2)$:
\begin{equation}
 z_n^m = \frac{\epsilon}{a^{\frac{1}{2}}}\sum_{j=0}^{n-1} \frac{\mu_j^m+\mu_{j+1}^m}{2},
\quad z_0^m=0.\label{eqn:z}
\end{equation}
Consequently, $(z_n^m, \theta_n^m)$ gives the discrete value of $\theta(z,t)$. It is remarkable
that, as a numerical scheme, this model can be regarded as a self-adaptive moving mesh scheme
\cite{FMO:CH,FMO:SP1,FMO:SP2}, since the step size in space is approximately given by
\begin{equation}
 z_{n+1}^m - z_n^m =  \frac{\epsilon}{a^{\frac{1}{2}}}\,\mu_n^m + O(\epsilon^2) 
= \frac{\epsilon a^{\frac{1}{2}}}{b-\theta_n^m} + O(\epsilon^2).
\end{equation}
Actually the grid points are dense for small $\theta$ and become sparse as $\theta$ increases. This
is likely to yield benefits at early times when the surface water content is still low, but
increases extremely rapidly due to the applied surface flux.

\begin{figure}[ht]
\begin{tabular}{ll}
\includegraphics[scale=\mypicscale]{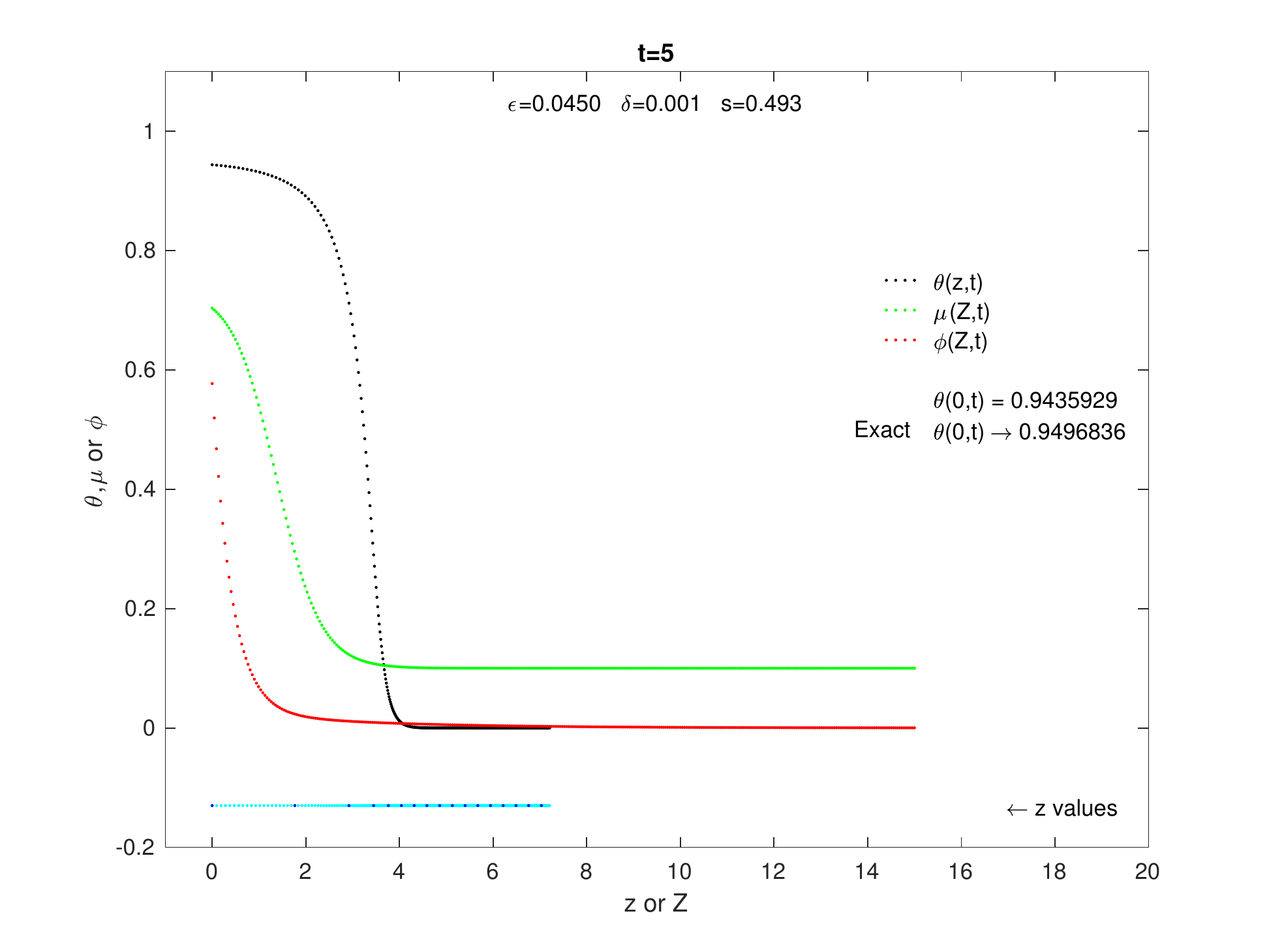}& \hspace{\mypichin}
\includegraphics[scale=\mypicscale]{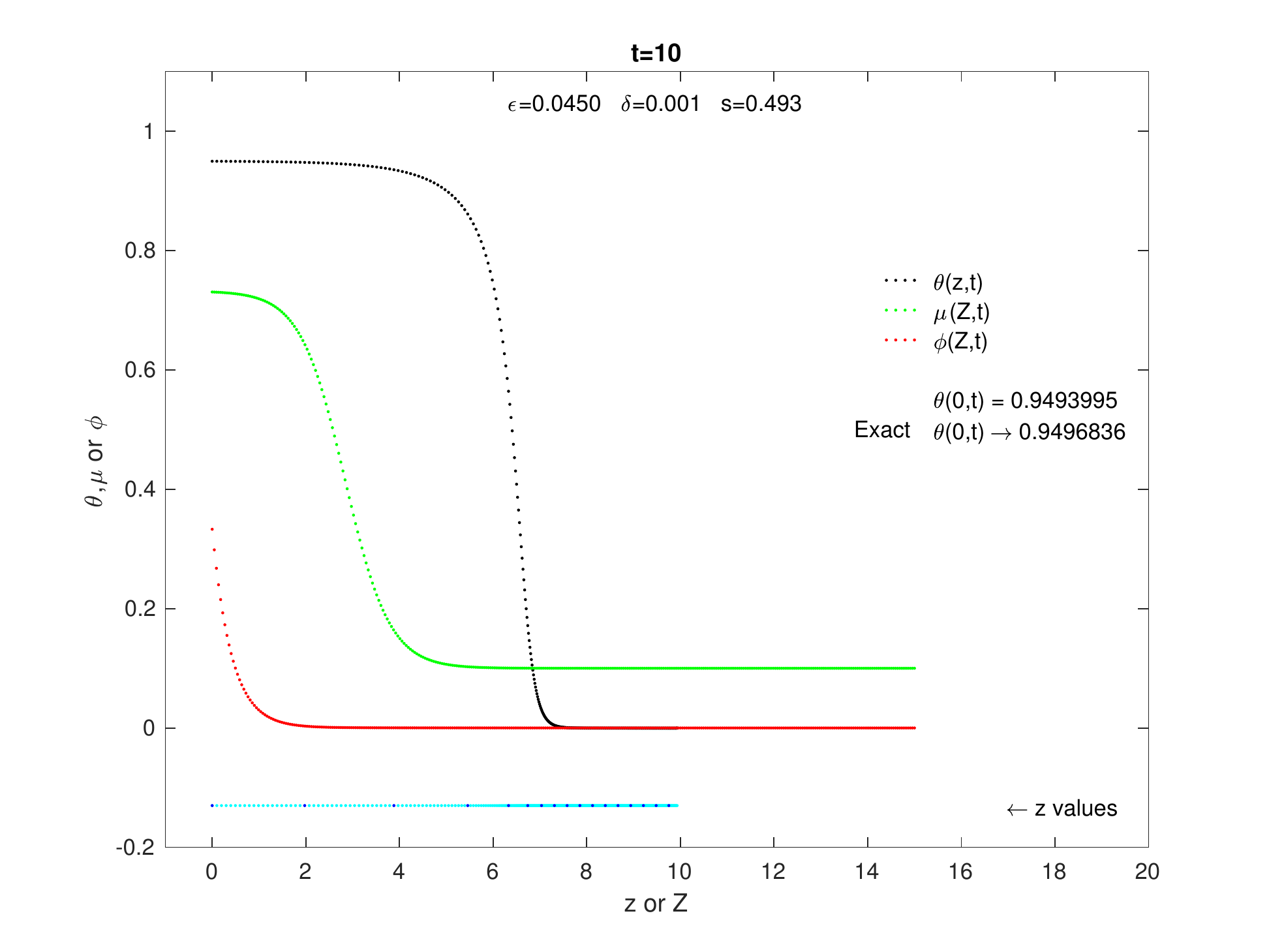} \vspace{\mypicvin}\\
\includegraphics[scale=\mypicscale]{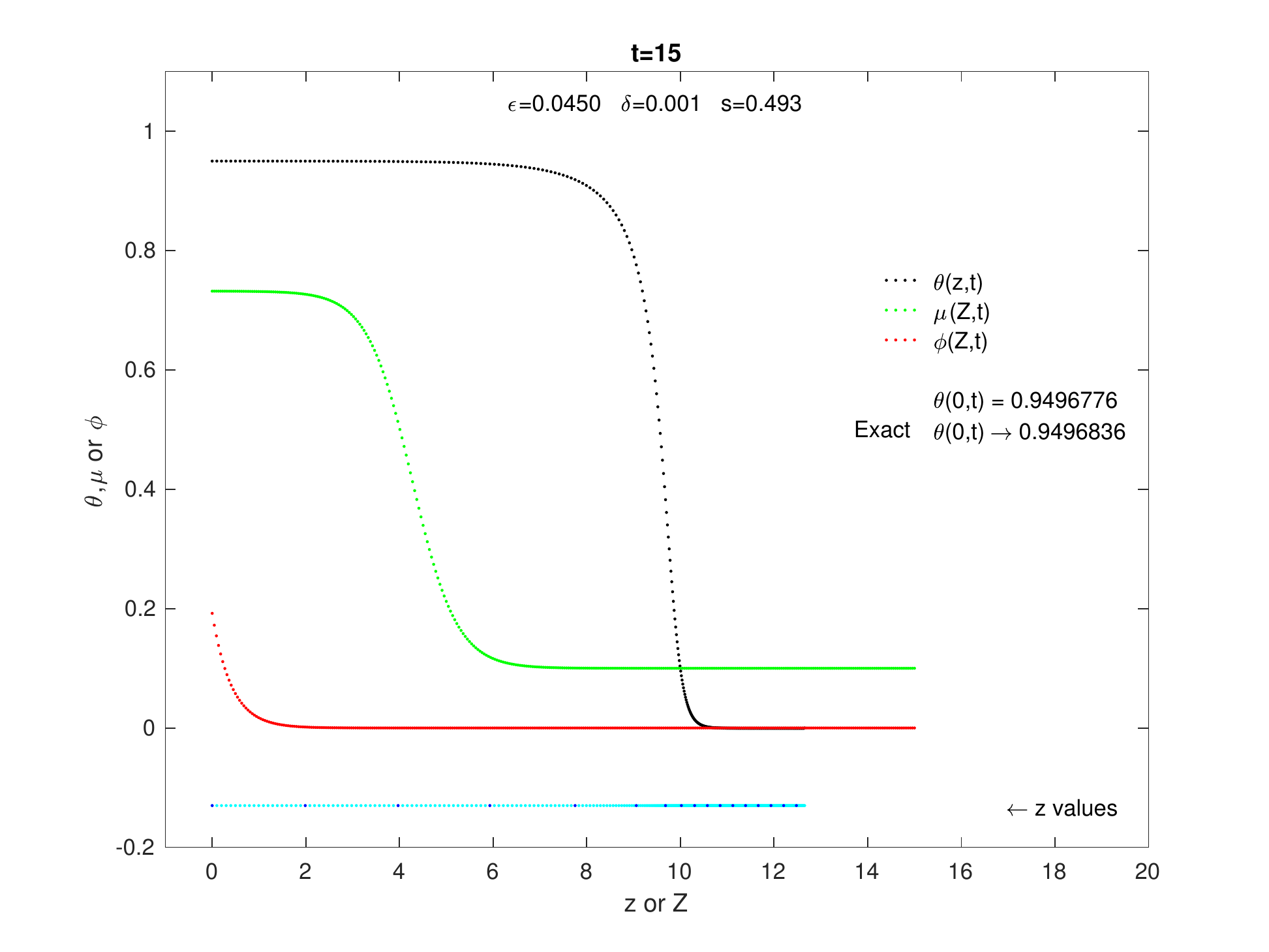} & \hspace{\mypichin} 
\includegraphics[scale=\mypicscale]{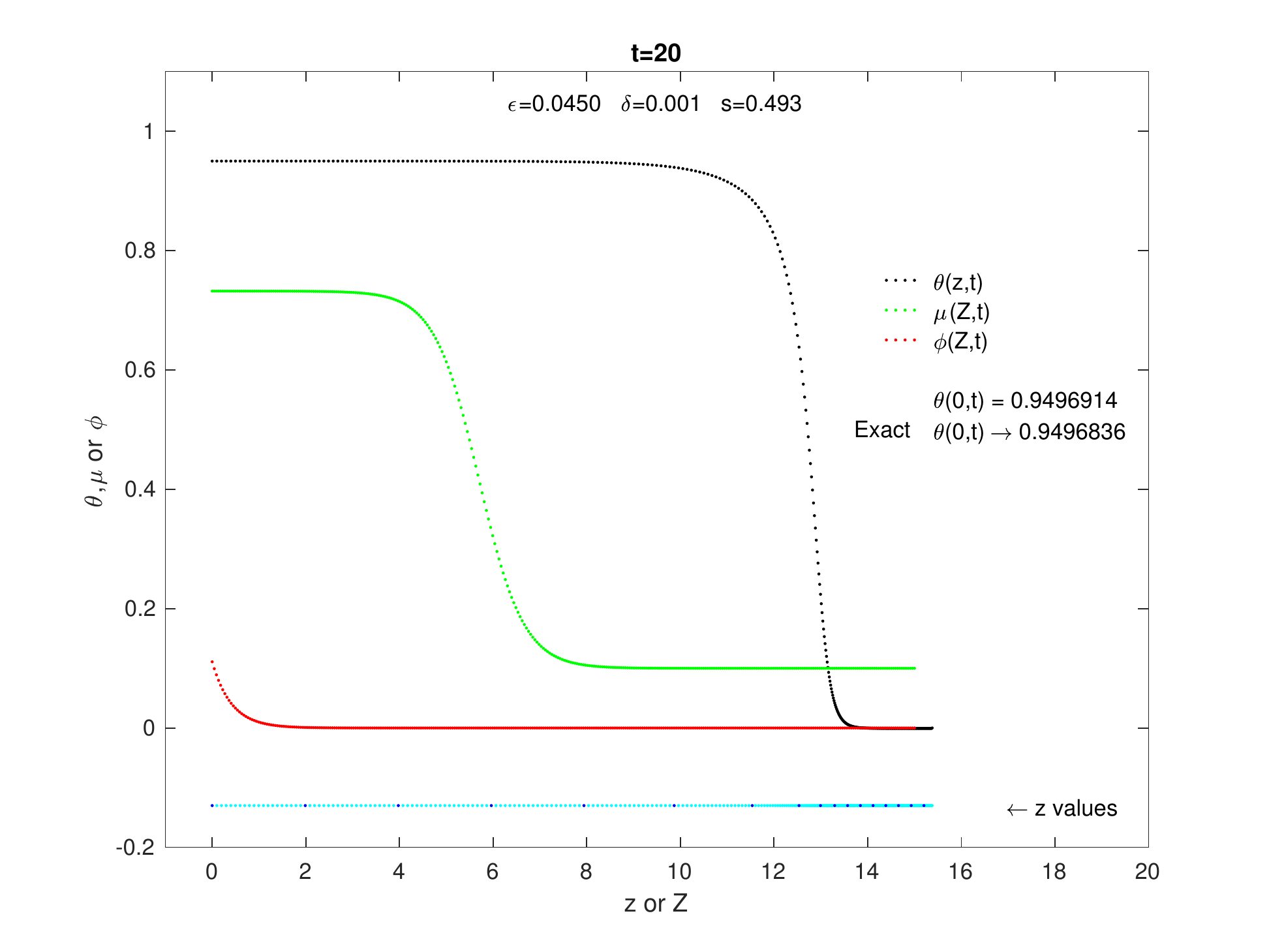}
\end{tabular}
\caption{Simulation of the linear model based on \eqref{eqn:dBurgers1} with constant surface flux $R(t)=0.6$ and 
the initial value $\theta(z,0)=0$. Parameters are given by $C=1.1$, $\epsilon=0.045$, $\delta=0.001$ and
$s=\frac{\delta}{\epsilon^2}=0.493$.} \label{fig:Euler_s0493}
\end{figure}

\begin{figure}[ht]
\begin{tabular}{ll}
\includegraphics[scale=\mypicscale]{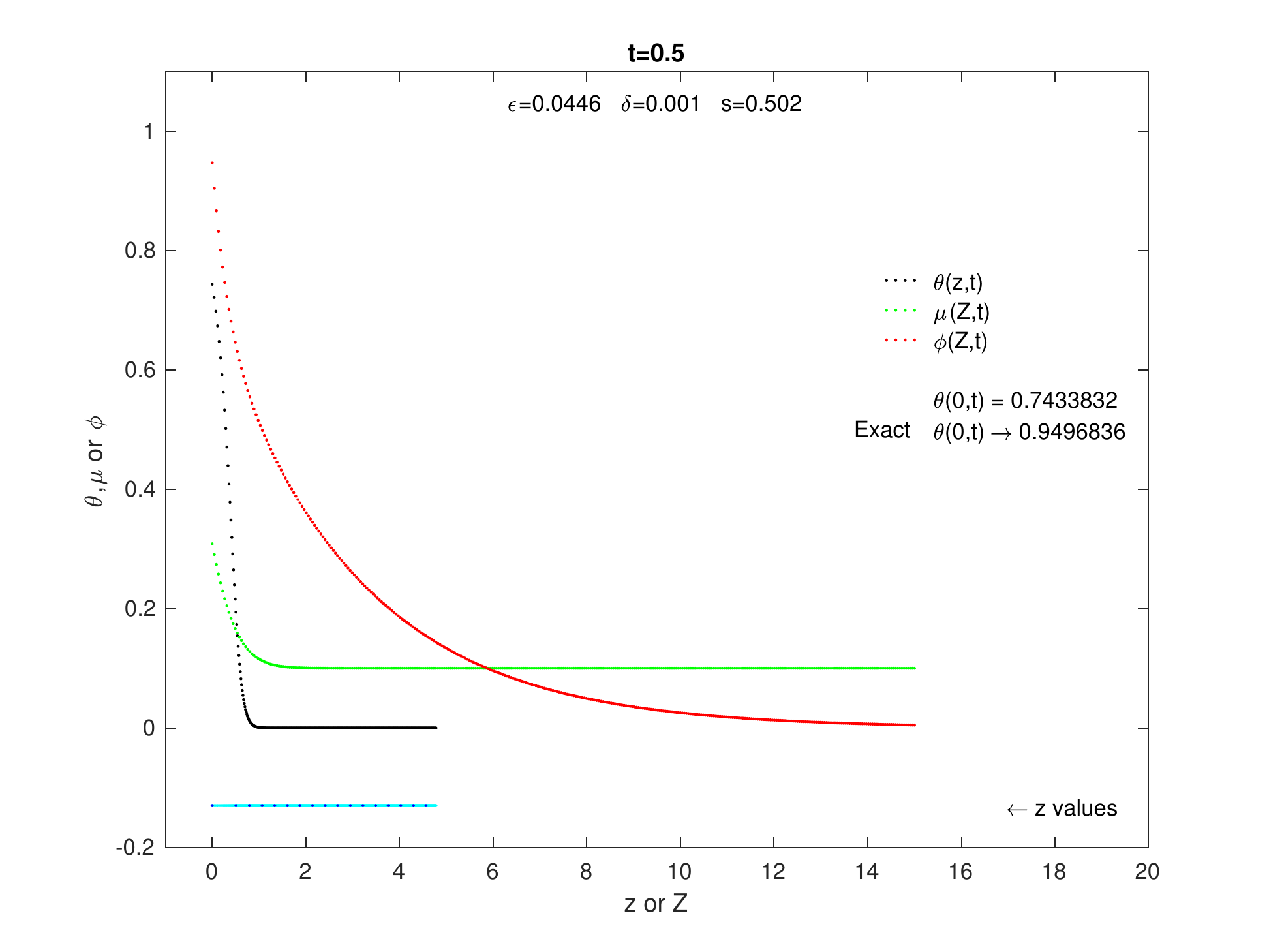}  &  \hspace{\mypichin}
\includegraphics[scale=\mypicscale]{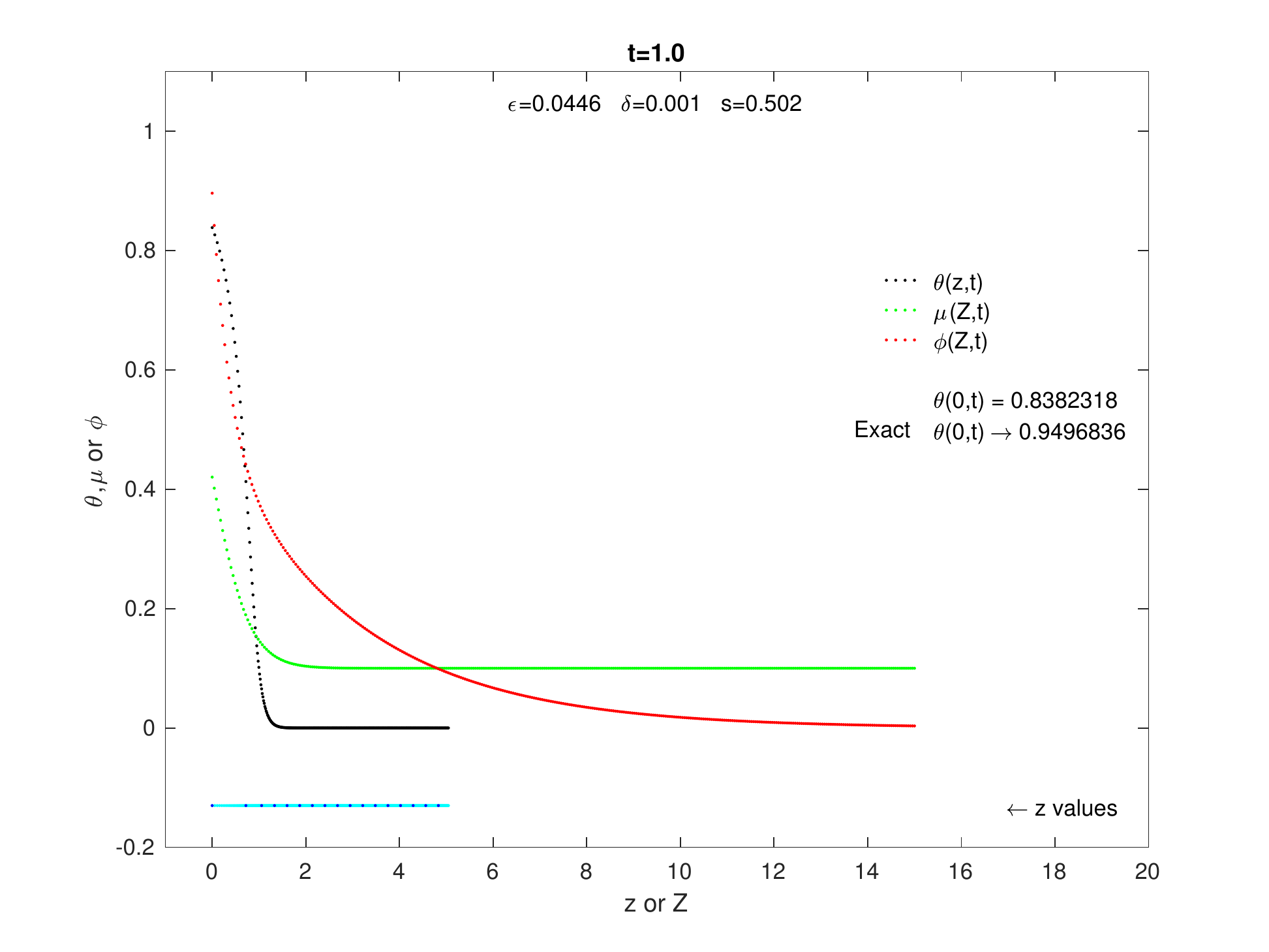} \vspace{\mypicvin}\\
\includegraphics[scale=\mypicscale]{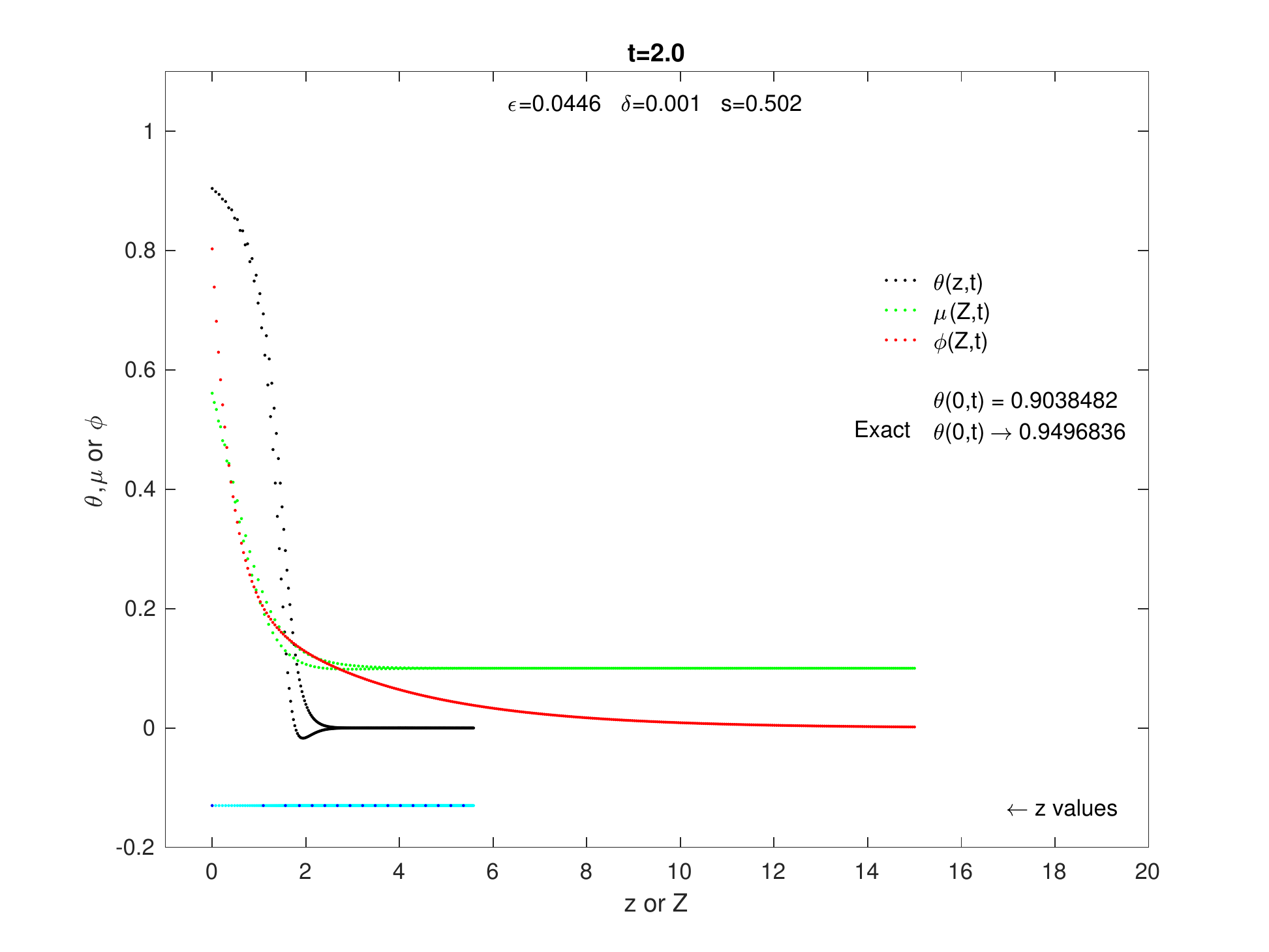} &  \hspace{\mypichin}
\includegraphics[scale=\mypicscale]{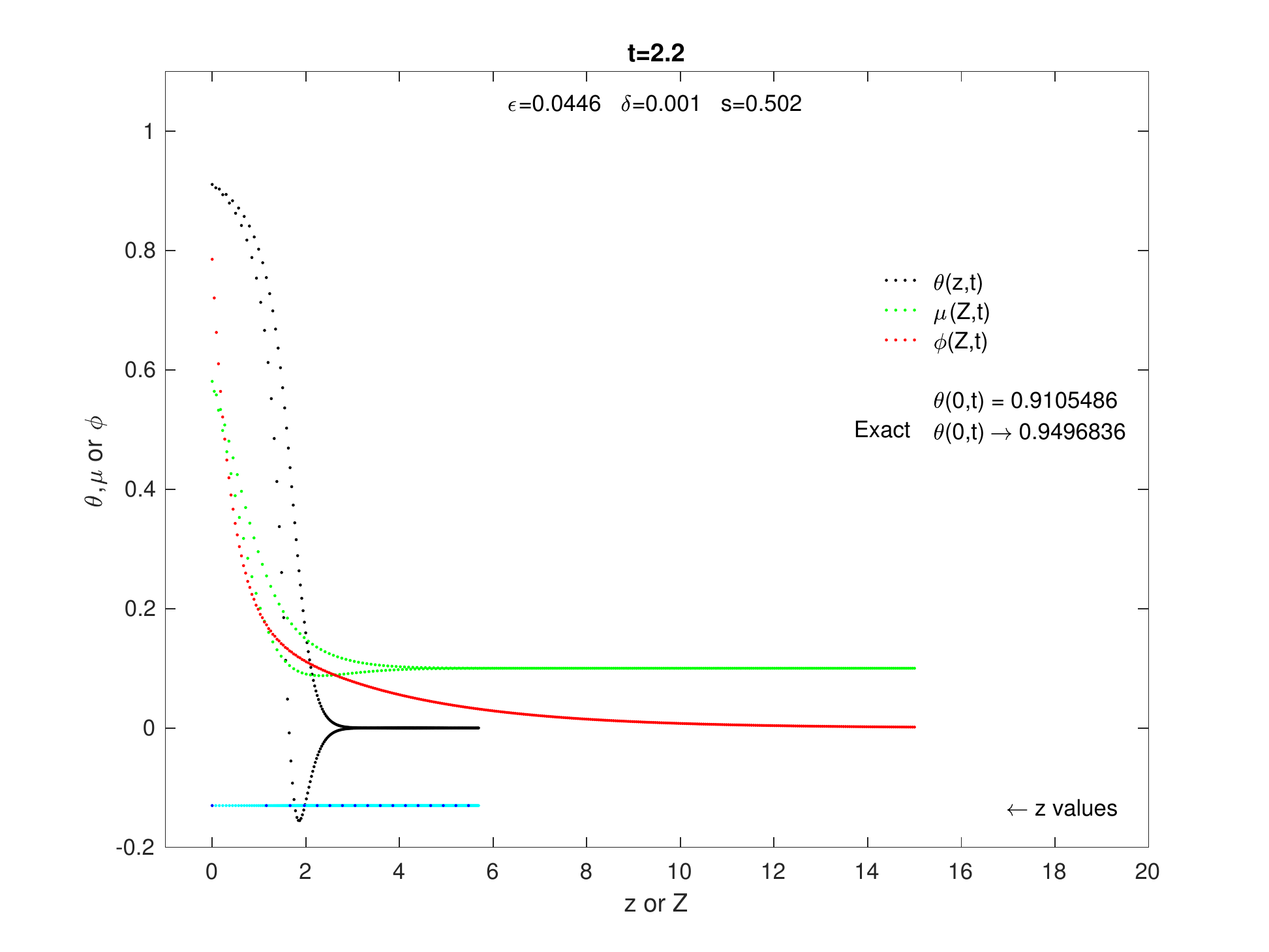}
\end{tabular}
\caption{Simulation of the linear model with the same conditions as Figure \ref{fig:Euler_s0493},
but with lattice intervals $\epsilon=0.0446$, $\delta=0.001$ and
$s=\frac{\delta}{\epsilon^2}=0.502$. Oscillation due to numerical instability starts around
$t=2.0$.} \label{fig:Euler_s0502}
\end{figure}

In summary, the integrable linear model can be computed as follows:
\begin{enumerate}
 \item Give the initial value $\phi_n^0$ for $n=1,2,\ldots,N$ by \eqref{eqn:discrete_model_phi_ic_sol}.
 \item For $m=1,2\ldots$ compute the following.
\begin{enumerate}
 \item Determine $\phi_1^m$ from
       \eqref{eqn:discrete_model_phi_bc_sol}, then 
compute $\phi_n^m$ for $n=2,\ldots,N-1$ using \eqref{eqn:dBurgers1} given $\phi_n^{m-1}$ ($n=1,\ldots,N$).
       \vspace{0.1cm}
 \item Compute $\phi_N^m$ by $\phi_N^m = P_{N-1}\phi_{N-1}^m$.
   \vspace{0.1cm}
 \item Compute $\mu_n^m$ and $\theta_n^m$ for $n=1,\ldots,N-1$ 
by \eqref{eqn:discrete_Cole-Hopf_transformation} and \eqref{eqn:discrete_model_theta}, respectively.
\vspace{0.1cm}
 \item Compute $\mu_0^m$ and $\theta_0^m$ from \eqref{eqn:u}, \eqref{eqn:u0} and \eqref{eqn:mu0_by_u0}.
 \vspace{0.1cm}
 \item Compute $z_n^m$ by \eqref{eqn:z} for $n=0,1,\ldots, N-1$.
 \vspace{0.1cm}
 \item Plot $(z_n^m, \theta_n^m)$ for $n=0,1,\ldots,N-1$.
\end{enumerate}
\end{enumerate}
\begin{rem} \hfill
\begin{enumerate}
\item
As $u^m_n = 1- O(\epsilon)$, in practical numerical computation, storing values $1-u^m_n$ rather
than $u^m_n$ should be less conductive to loss of numerical precision.
\item
 The discrete counterpart of the flux $J$ may be introduced in terms of $\phi_n^m$ as
\begin{equation}
   J_n^m = -\frac{2a^{\frac{1}{2}}\epsilon}{\phi_{n+2}^m - \phi_n^m}
\left[
 \frac{\phi_{n+2}^{m} - 2\phi_{n+1}^{m} + \phi_n^m}{\epsilon^2} + a\phi_{n+1}^m 
- \frac{\beta}{a^{\frac{1}{2}}}\,\frac{\phi_{n+2}^m - \phi_n^m}{2\epsilon}
\right],
\end{equation}
which is an analogue of \eqref{eqn:continuous_flux_phi}, so that the condition $J_0^m=R^m$ yields
\eqref{eqn:discrete_model_phi_bc_eq1}. $J_n^m$ may be expressed in terms of $\mu_n^m$ or
$\theta_n^m$ by using \eqref{eqn:discrete_Cole-Hopf_transformation} and
\eqref{eqn:discrete_model_theta}, but we omit the concrete expression since it is complicated.
\end{enumerate}
\end{rem}
Figure \ref{fig:Euler_s0493} shows the numerical result starting from the initial value
$\theta(z,0)=0$ with constant surface flux $R(t)=0.6$ and $C=1.1$. In this case it is known that
$\lim\limits_{t \to \infty} \theta(0,t) = 0.94968353$ \cite{BroadbridgeWhite1988a}. Then taking
$\epsilon=0.045$ and $\delta=0.001$, we have $\theta^m_0|_{t=20}=0.9496914$ so that the precision is
$10^{-4}$. The self-adaptive nature of our numerical scheme is highlighted by plotting just the
$z$-values of node points at the bottom of each subplot, with every twentieth $z$-value coloured
darker blue.  We could choose smaller $\epsilon$ for improved accuracy, however, the linear
difference equation \eqref{eqn:dBurgers1} is a well-known example which causes numerical instability
according to the value of $s=\frac{\delta}{\epsilon^2}$; it is unstable when $s>\frac{1}{2}$. Figure
\ref{fig:Euler_s0502} shows the simulation with the same condition as Figure \ref{fig:Euler_s0493}
with lattice intervals $\epsilon=0.0446$, $\delta=0.001$ and $s=0.502>\frac{1}{2}$.  Oscillation due
to numerical instability occurs around $t=2.0$ and the calculation quickly crashes.  The restriction
$s<\frac{1}{2}$ makes accurate numerical simulation prohibitively difficult.

\begin{figure}[ht]
\begin{tabular}{ll}
\includegraphics[scale=\mypicscale]{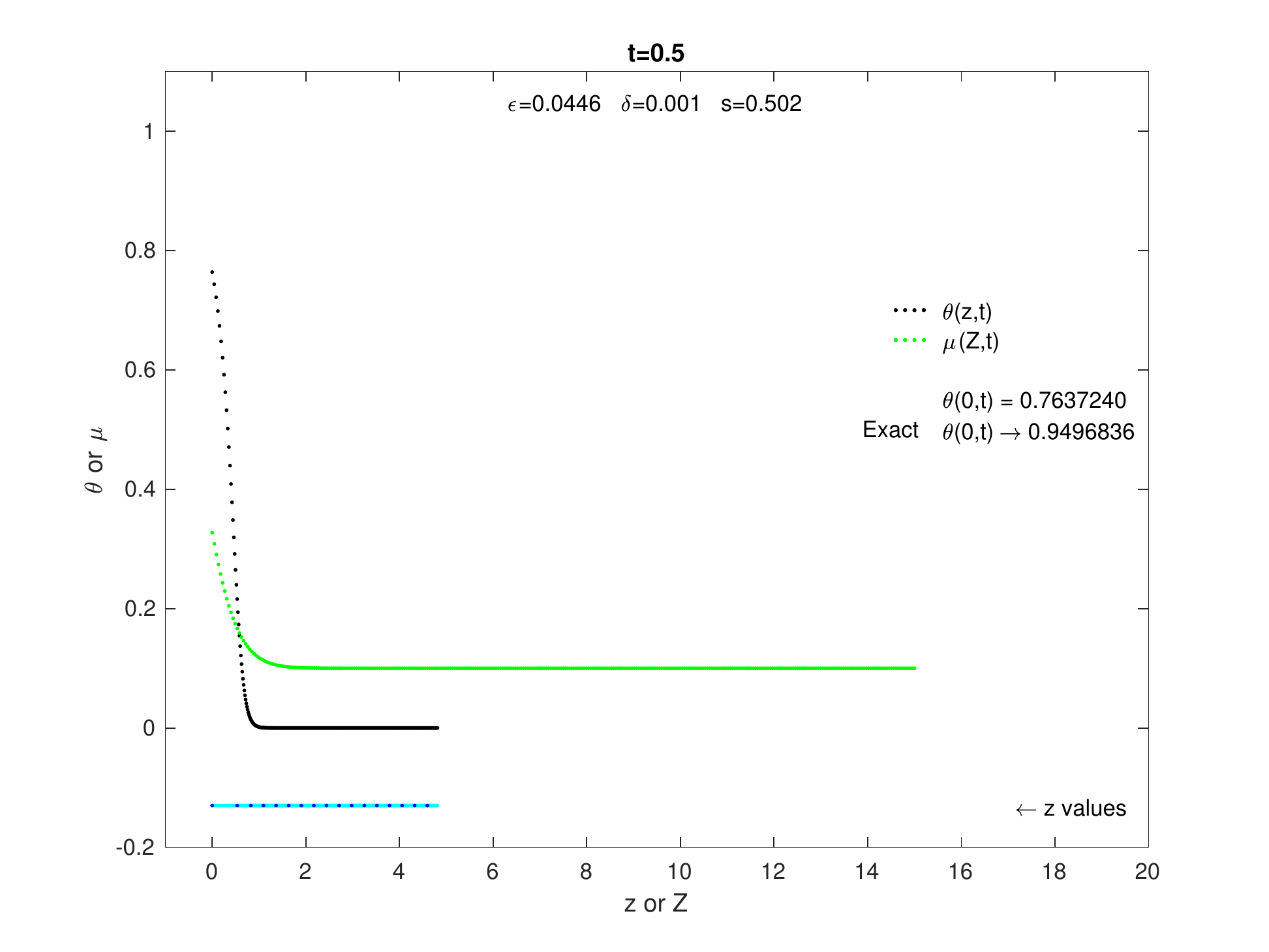}  & \hspace{\mypichin}
\includegraphics[scale=\mypicscale]{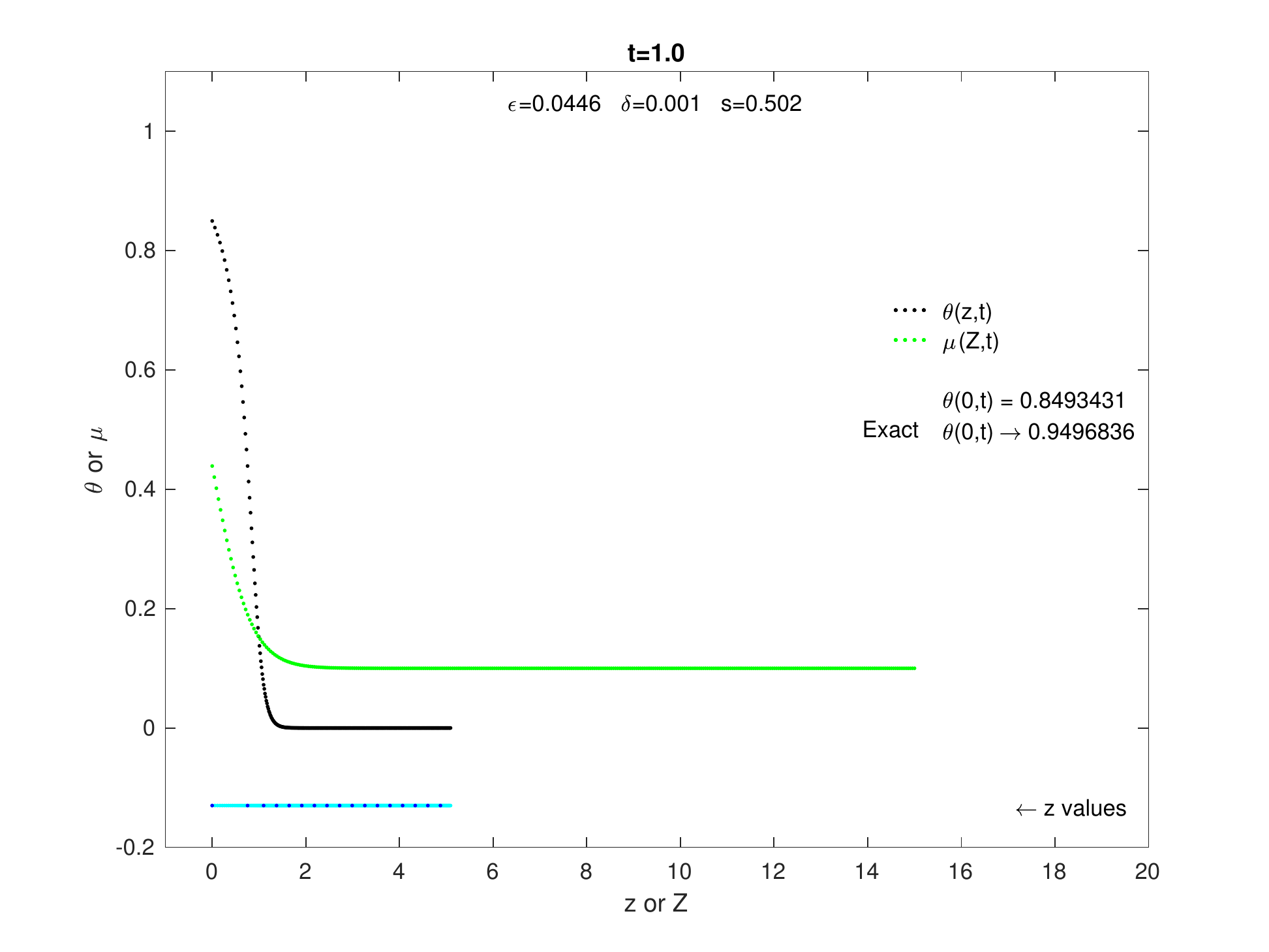} \vspace{\mypicvin}\\
\includegraphics[scale=\mypicscale]{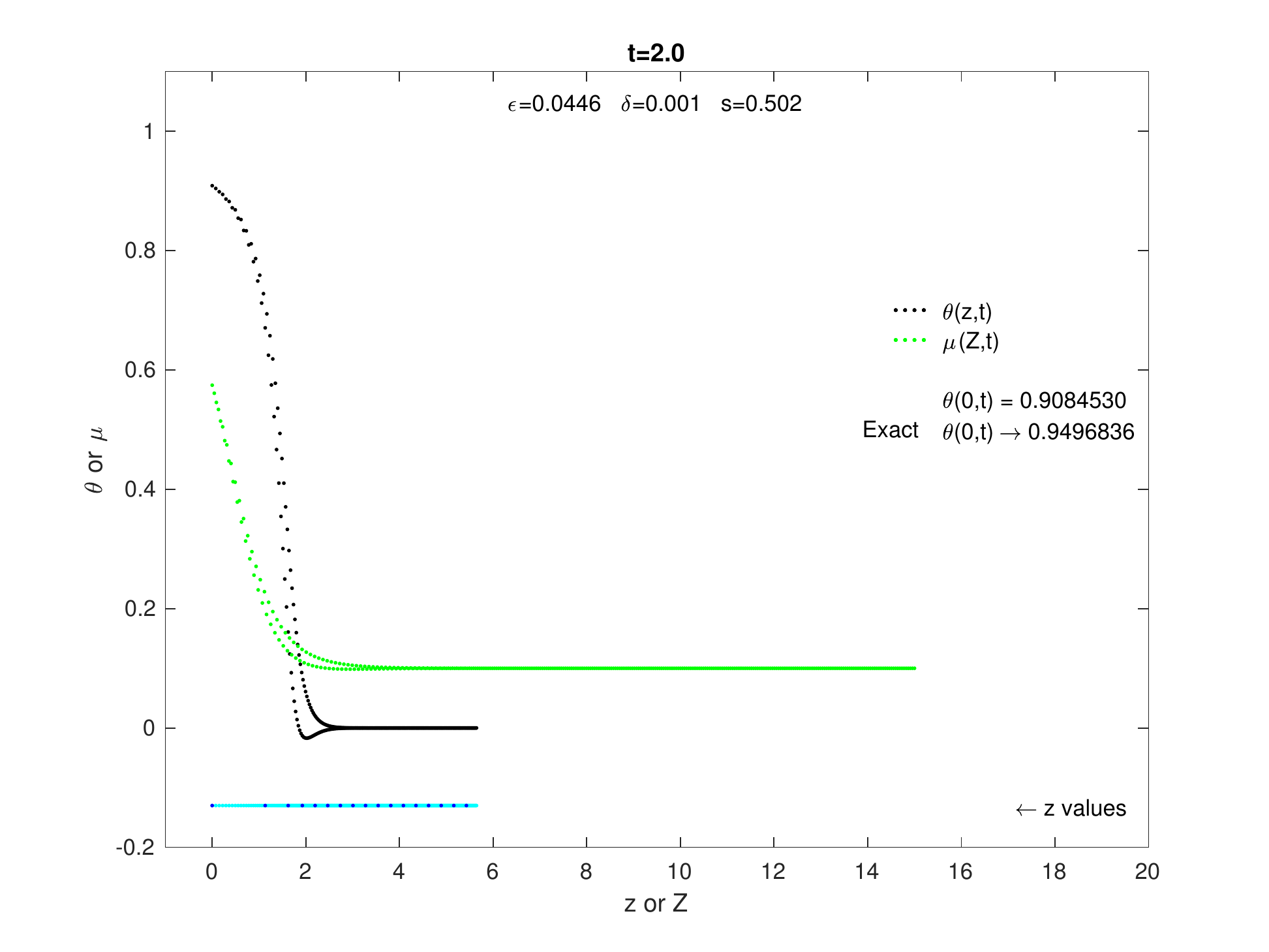} & \hspace{\mypichin}
\includegraphics[scale=\mypicscale]{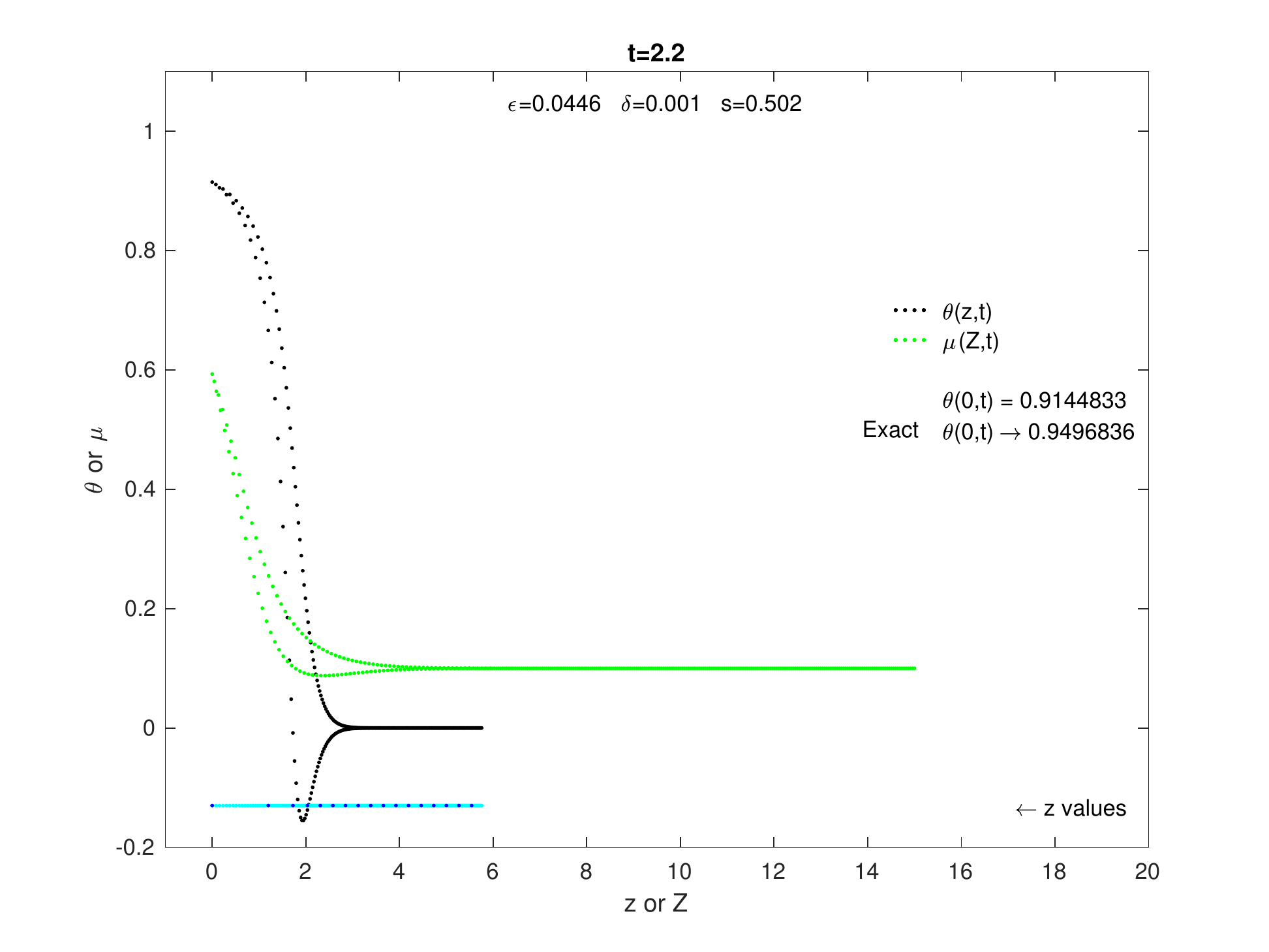}
\end{tabular}
\caption{Simulation of the discrete Burgers model under the same conditions as Figure \ref{fig:Euler_s0502}.
The numerical instability is inherited from the linear model.} \label{fig:Euler_mu_s0502}
\end{figure}

The numerical instability for \eqref{eqn:dBurgers1} is a consequence of linear stability analysis.
So one might think that we could avoid the instability by adopting the nonlinearized scheme, namely
the discrete analogue of the Burgers equation. To this end, it is convenient to write down the scheme in
terms of $u_n^m$ \eqref{eqn:u}. We then have the discrete Burgers equation
\cite{HerederoLeviWinternitz1999,Hirota:difference5,Nishinari_Takahashi:Burgers}
\begin{equation}
\begin{split}
& \frac{u^{m+1}_{n}}{u^m_{n}} 
= \frac{1 + \frac{\delta}{\epsilon^2}\left[u_{n+1}^m - 2  + \frac{1}{u_{n}^m}
+  \kappa^m\, \left(u_{n+1}^m - \frac{1}{u_{n}^m} \right)\right] }
{1 + \frac{\delta}{\epsilon^2}\left[u_n^m - 2  + \frac{1}{u_{n-1}^m}
+  \kappa^m\, \left(u_n^m - \frac{1}{u_{n-1}^m} \right)\right]},\quad
\kappa^m = \frac{\epsilon(R^m - \beta)}{2a^{\frac{1}{2}} }
\\[2mm]
&\hspace*{40pt} n=1,\ldots,N-1, \quad m=0,1,2,\ldots,
\end{split}
\label{eqn:dBurgers1_u}
\end{equation}
with initial condition 
\begin{equation}
u_n^0 = \frac{1-\frac{2a^{\frac{1}{2}}}{b\epsilon}\mu^{(0)}_n}{1+\frac{2a^{\frac{1}{2}}}{b\epsilon}\mu^{(0)}_n},\quad
n=0,2\ldots,N-1,
\label{eqn:discrete_model_u_ic_eq}
\end{equation}
and the boundary conditions
\begin{equation}
  u_0^m = \frac{1-\kappa^m}{2 - a\epsilon^2 - (1+\kappa^m)u_1^m},\quad u_{N}^m = P_{N-1}.
\end{equation}
Note that $\mu_n^m$ and $\theta_n^m$ are recovered by
\begin{equation}
 \mu_n^m =  -\frac{2a^{\frac{1}{2}}}{b\epsilon}\, 
\frac{u_{n}^m-1}{u_{n}^m + 1},\quad
\theta_n^m = b\left(1 + \frac{a^{\frac{1}{2}}\epsilon}{2}\, 
\frac{u_{n}^m + 1}{u_{n}^m-1}\right).\label{eqn:mu_by_u}
\end{equation}
Then we plot $(z_n^m,\theta_n^m)$ with \eqref{eqn:z}. Figure \ref{fig:Euler_mu_s0502} illustrates
the numerical result under the same condition and parameters as Figure \ref{fig:Euler_s0502}.  This
gives the same result, and unfortunately the numerical instability is also inherited from the linear
model. Indeed, choosing the lattice intervals such that $s<\frac{1}{2}$, the numerical computation
is stable with sufficient precision for $\theta(0,t)$ at large $t$.

\subsection{A stable discrete integrable model: Crank--Nicolson scheme}\label{subsec:discrete_Crank-Nicolson}
In order to overcome the numerical instability, a simple alternative to \eqref{eqn:dBurgers1} with
second-order accuracy is the Crank--Nicolson (CN) Scheme:
\begin{align}
&2 \frac {\phi^{m+1}_{n} - \phi^m_{n}}{\delta} = F^m_n + F^{m+1}_n, 
\nonumber \\
&F^m_n = \frac{\phi^{m}_{n+1} - 2 \phi^{m}_{n} + \phi^{m}_{n-1}}{\epsilon^2} 
+  \frac {R^{m} - \beta}{a^{\frac{1}{2}} }\, \frac{\phi^{m}_{n+1} - \phi^{m}_{n-1} }{2 \epsilon}.
\label{eqn:dBurgers3}
\end{align}

Choosing other procedures, such as the discrete Cole-Hopf transformation
\eqref{eqn:discrete_Cole-Hopf_transformation} and the hodograph transformation \eqref{eqn:z}, to be
the same as the previous case, it is possible to set the initial condition by
\eqref{eqn:discrete_model_phi_ic_sol} and the large-$Z$ boundary condition as $\phi_N^m =
P_{N-1}\phi_{N-1}^m$.

Our previous surface boundary condition \eqref{eqn:discrete_model_phi_bc_eq1} was not accurate to
second-order, and hence requires modification. Our Cole--Hopf transformation
\eqref{eqn:discrete_Cole-Hopf_transformation} is accurate to second-order when centered on the
half-node position $n+1/2$, as discussed in Remark \ref{etaremark}. Accordingly, the precise place
to apply the surface boundary condition when considering $\phi$ is the half node position $n =1/2$,
as this corresponds to $\mu^m_0$ and $\theta^m_0$ to second-order accuracy in $\epsilon$:
\begin{align}
\frac 2 \delta \big(\phi^{m+1}_{1/2} - \phi^{m}_{1/2}\big) = -a \big( \phi^{m+1}_{1/2} + \phi^{m}_{1/2}  \big),
 \quad 
\phi^{m}_{1/2} = \frac 12 \big( \phi^{m}_{0} + \phi^{m}_{1}  \big).
\end{align}
Here the averaging of nodes $\phi^{m+1}_{0}$ and $\phi^{m}_{0}$ on the right-hand-side of the
equation ensures that the boundary condition also is of second order accuracy in $\delta$ when
centered on the half node position $m+ 1/2$, similar to the CN scheme itself.  This surface boundary
condition can be alternatively expressed as
\begin{equation}
\phi^{m+1}_{0} + \phi^{m+1}_{1} = \frac {2-a \delta}{2+ a \delta} \big( \phi^{m}_{0} + \phi^{m}_{1} \big).
\label{eqn:intCN_surfaceBC}
\end{equation}

In order
to compute $\phi_n^m$ ($n=0,\ldots,N-1$) we solve the following system of linear equations:
\begin{align}
&A\Phi = B,
\label{eq:linsys} \\
&\Phi = \left[\phi_0^{m+1},\phi_1^{m+1},\ldots,\phi_{N-1}^{m+1}\right]^{T},
\nonumber \\
&B=\left[B_0, B_1,\ldots,B_{n-1}\right]^{T},
\nonumber \\
&A \!=\!\!
\left[\!\!
\begin{array}{cccccc}
A_{0,0}	& A_{0,1}		&   0		 	&   {\s 0}		&   {\s \cdots}	&   {\s 0} \\
A_{1,0}	& A_{1,1}   	&   A_{1,2}  	&   {\s 0}  		&   {\s \cdots}	&   {\s 0} \\
{\s  0 }       & A_{2,1} 		&   A_{2,2}  	&    A_{2,3}	&   {\s \cdots}	&   {\s 0} \\
{\s \vdots}	&  {\s \ddots}	& {\s \ddots}	& {\s \ddots}	&   {\s \ddots}    & {\s \vdots}\\
{\s 0}		&  {\s \cdots}     & A_{N-3,N-4}	& A_{N-3,N-3}	&A_{N-3,N-2} 		&   {\s 0} \\
{\s 0} 	&  {\s \cdots} 	&   {\s 0} 		&A_{N-2,N-3}	& A_{N-2,N-2} 		& A_{N-2,N-1} \\
{\s 0} 	&  {\s \cdots} 	&   {\s 0} 		&   {\s 0} 		&A_{N-1,N-2}  		& A_{N-1,N-1}
\end{array}
\!\!\right]\!\!.
\nonumber
\end{align}
Here $A_{0,0} = A_{0,1} =1$ are obtained directly from \eqref{eqn:intCN_surfaceBC}, so that $B_0$ is given by the right-hand-side of the equation. From \eqref{eqn:dBurgers3} we have for $ k = 1, 2, \ldots, N-2$:
\begin{align}
&A_{k,k-1} = -s(1-\kappa^{m+1}), \;\;
A_{k,k} = 2(1+s), \;\;
A_{k,k+1} = -s(1+\kappa^{m+1}),
\nonumber \\
&B_k = s(1-\kappa^{m})\phi^m_{k-1} +  2(1-s)\phi^m_k + s(1+\kappa^{m})\phi^m_{k+1}.
\end{align}
Finally, incorporating the boundary condition $\phi_N^m = P_{N-1}\phi_{N-1}^m$ gives
\begin{align}
&A_{N-1,N-2} = -s(1-\kappa^{m+1}), \;\;
A_{N-1,N-1} = 2(1+s)-s P_{N-1}(1+\kappa^{m+1}) ,
\nonumber \\
&B_{N-1} = s(1-\kappa^{m})\phi^m_{N-2} +  \big[2(1-s) + s P_{N-1} (1+\kappa^{m})\big] \phi^m_{N-1}.
\end{align}
Hence we obtain $\phi_n^m$ for $n=0,1,\ldots, N$ at each $m$.  

\begin{figure}[ht]
\begin{tabular}{ll}
\includegraphics[scale=\mypicscale]{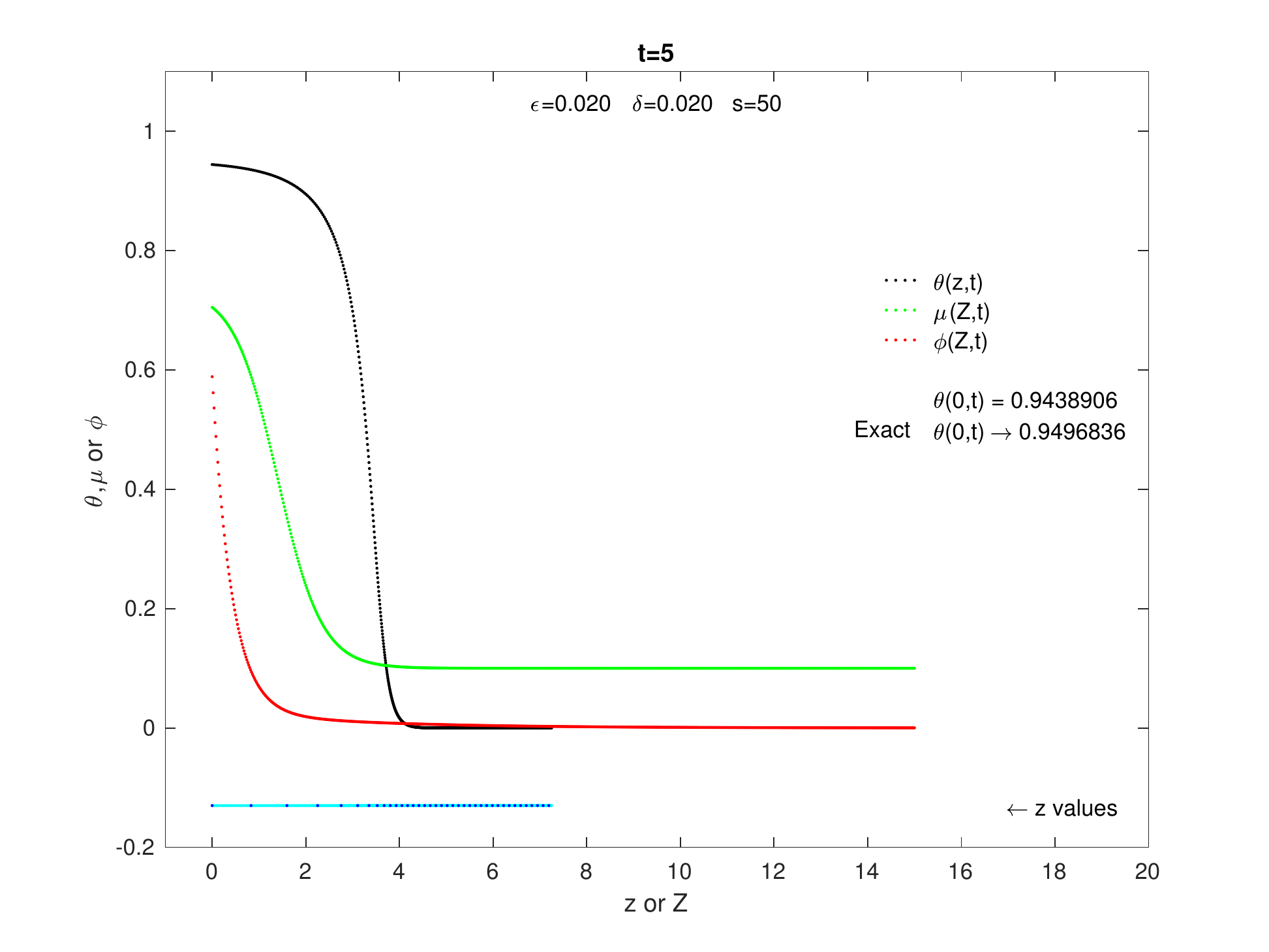}  & \hspace{\mypichin}
\includegraphics[scale=\mypicscale]{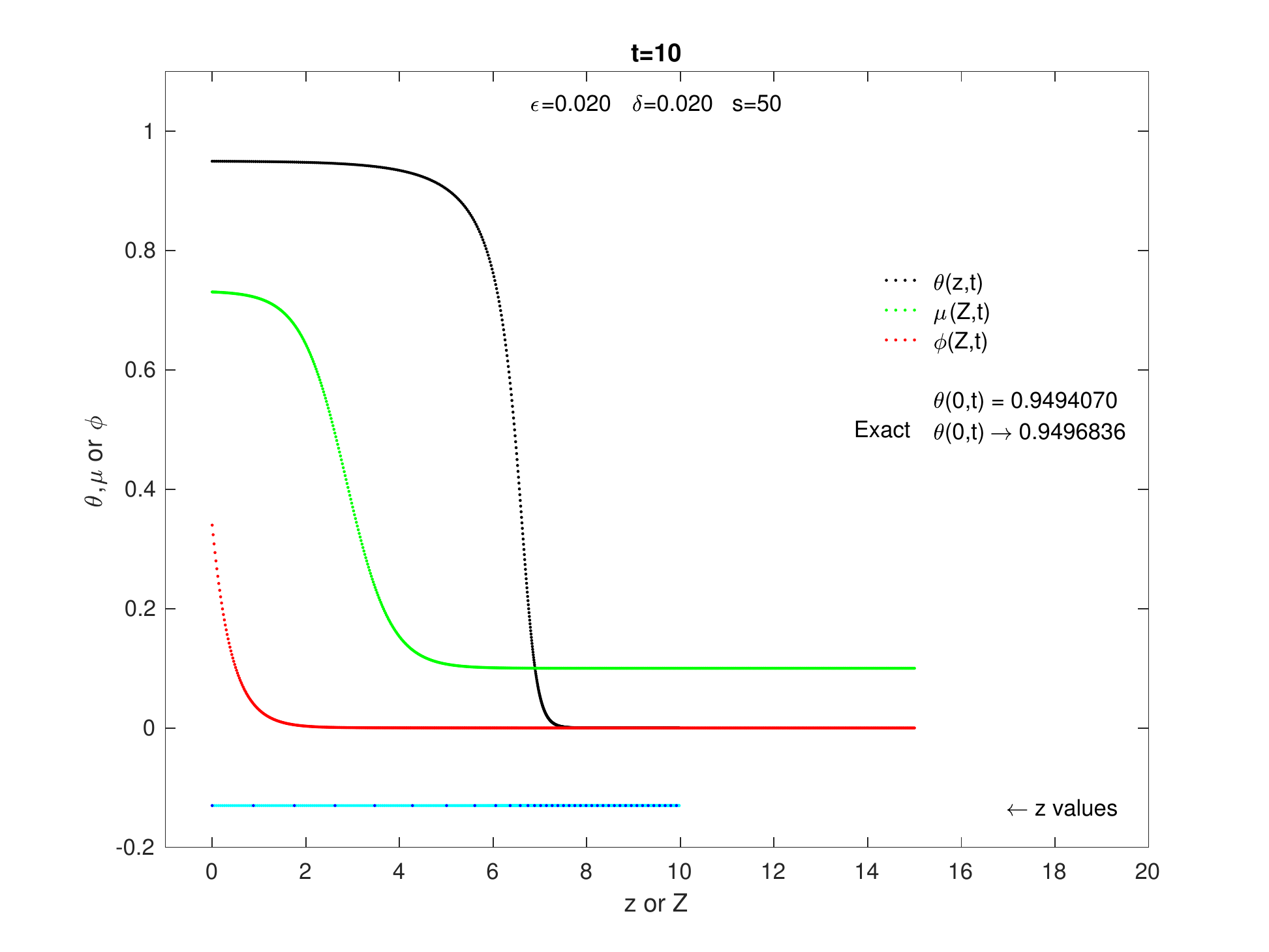} \vspace{\mypicvin}\\
\includegraphics[scale=\mypicscale]{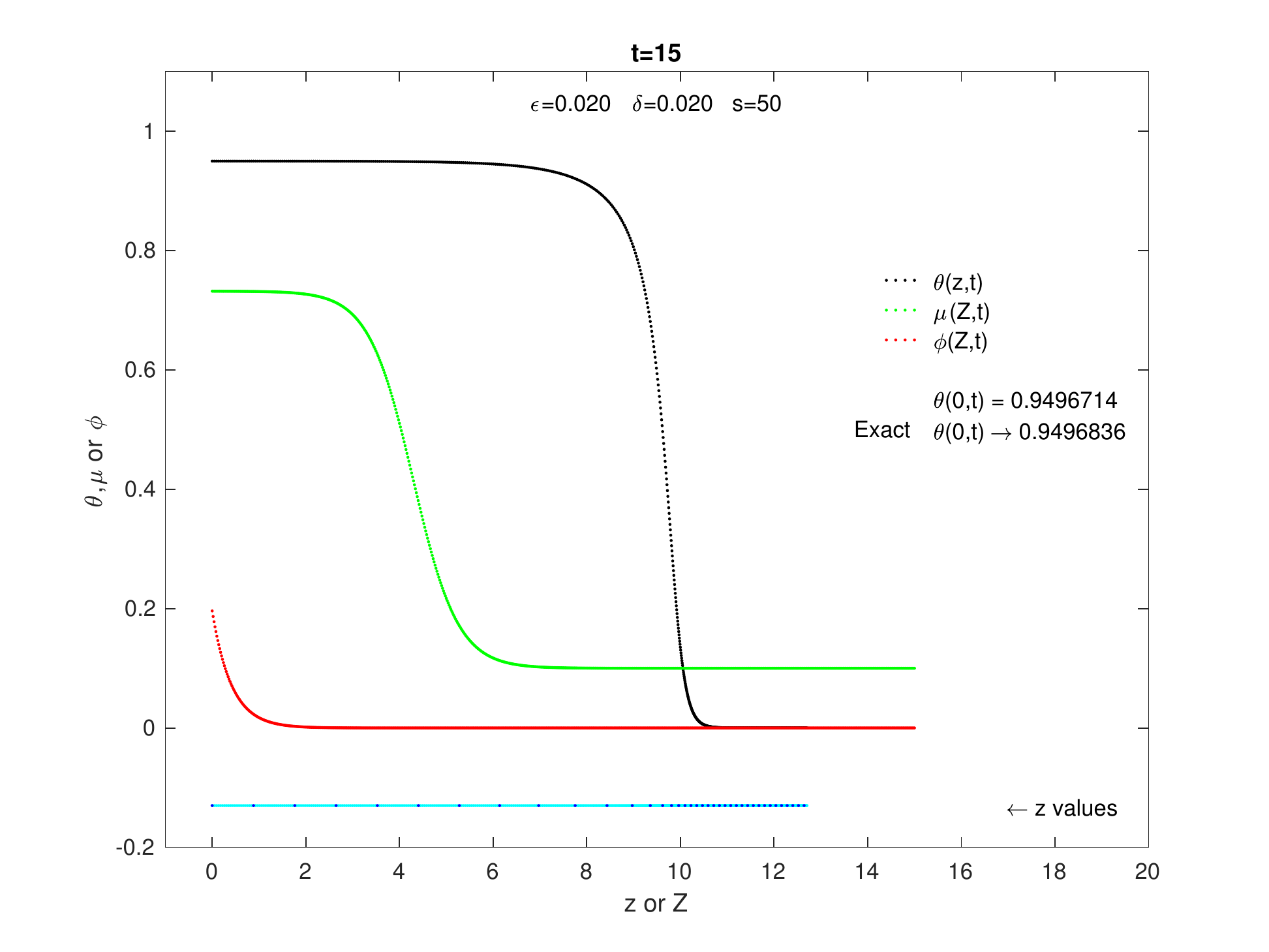} & \hspace{\mypichin}
\includegraphics[scale=\mypicscale]{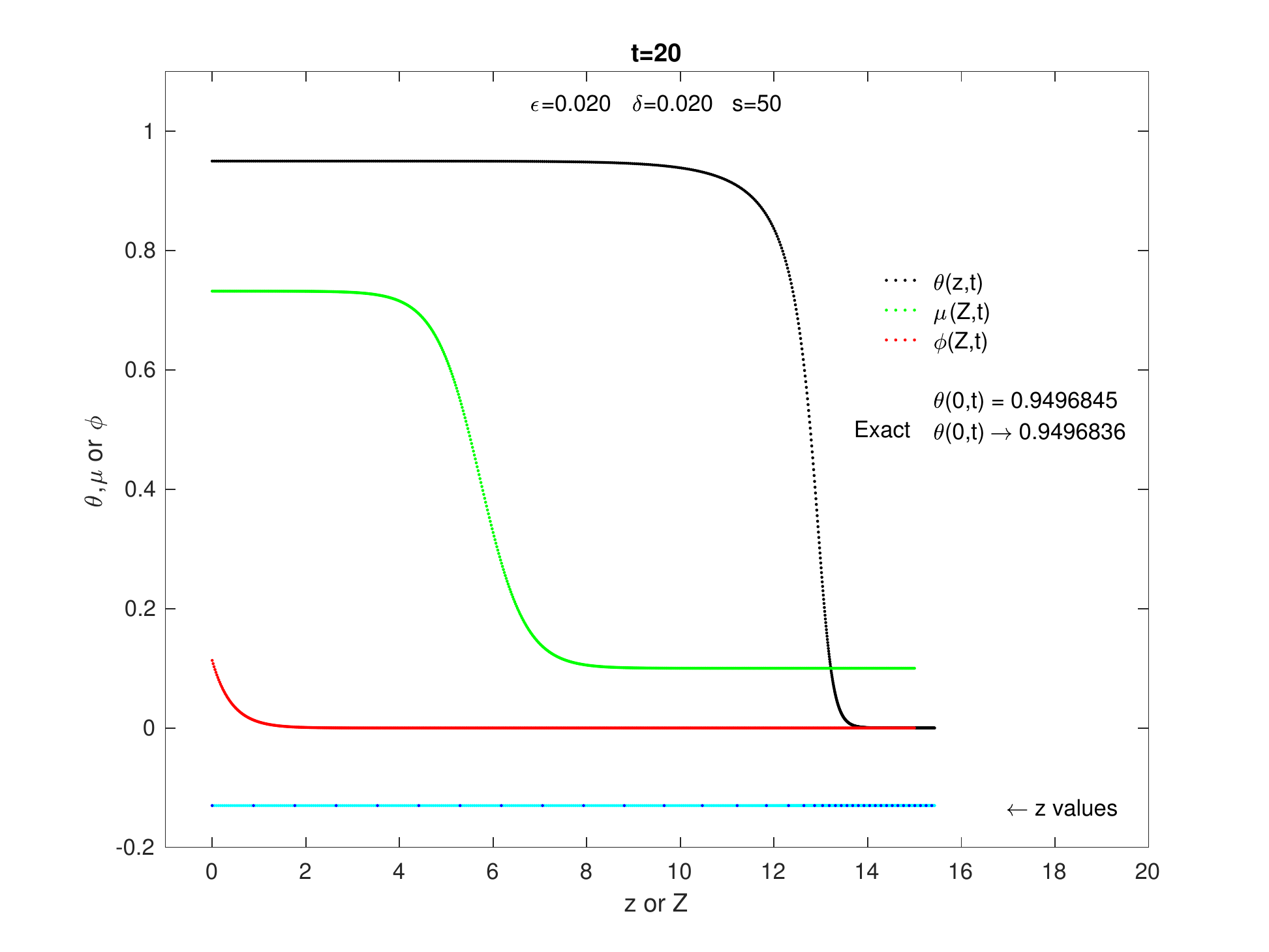}
\end{tabular}
\caption{Simulation of the linear model based on the Crank--Nicolson scheme under the same
conditions as Figure \ref{fig:Euler_s0493}, but with lattice intervals $\epsilon=\delta=0.02$,
$s=50$. The computation is stable and the computed value of $\theta(0,t)$ is accurate to $10^{-5}$.
} \label{fig:CrankNicolson_s50}
\end{figure}

Numerical results obtained are almost identical to Figure \ref{fig:Euler_s0493} for the same
boundary conditions and parameters. Figure \ref{fig:CrankNicolson_s50} shows numerical results under
the same conditions, but with lattice intervals given by $\epsilon=\delta=0.02$, $s=50$. As
expected, computations are stable regardless of value of $s$.

Therefore, the discrete integrable model based on the Crank--Nicolson scheme may provide stable,
reasonably accurate calculations for modelling groundwater infiltration. 

\section{Comparison with direct computation}\label{subsec:direct_comparison}

The nonlinear equation \eqref{eqn:continuous_model_theta} is amenable to direct computational
solution via a variety of well established methods. If we put aside matters of theoretical interest
discussed above, we have essentially established the feasibility of the Crank--Nicolson integrable
model, but have not yet demonstrated any practical advantage over direct numerical approaches.
While numerical methods with higher-order accuracy might be adopted to solve
\eqref{eqn:continuous_model_theta}, they would not constitute a fair test in the present context: it
should be clear that the integrable model has the potential to be generalised such that higher-order
accuracy is achieved, but that is beyond the natural scope of this article. As such, in order to
assess the numerical performance of our integrable model we will proceed to implement the
Crank-Nicolson scheme directly on the original equation \eqref{eqn:continuous_model_theta} for
$\theta$:
\begin{align}
&2 \frac {\theta^{m+1}_{n} - \theta^m_{n}}{\delta} = F^m_n + F^{m+1}_n, 
\nonumber \\
&F^m_n = \left[ \frac ab - \frac{ ab}{(b-\theta^m_n)^2}\right] \frac{\theta^{m}_{n+1} - \theta^{m}_{n-1}}{2\epsilon} 
\nonumber \\
& \qquad + 
\frac a {\epsilon^2} 
\frac{\theta^{m}_{n+1} - \theta^{m}_{n}}{(b-\theta^m_{n+1})(b-\theta^m_n)} -\frac{\theta^{m}_{n} - \theta^{m}_{n-1}}{(b-\theta^m_{n})(b-\theta^m_{n-1})}.
\label{eqn:NL_CN}
\end{align}
Note that in this context the lattice parameters now relate to $z$ and $t$, so that $\theta(z,t) =
\theta(n \epsilon,m \delta) = \theta^m_n$.  Our initial condition is $\theta^0_n =0$ for $n = 1,
\ldots, N$, and under reasonable assumptions already discussed, we can assume that $\theta^m_N =
0$. Our flux boundary condition at $z=0$ was given in \eqref{eqn:continuous_model_theta_ic_bc}:
\begin{equation}
(b - \theta)(R(t)-\beta) = a b + \frac a b (b -\theta)^2 - \frac a {(b-\theta)}\frac {\partial \theta}{\partial z}.
\label{eqn:continuous_theta_ic_again}
\end{equation}
We have discretized spatial derivatives in our integrable model both to define the Cole--Hopf
transformation \eqref{eqn:discrete_Cole-Hopf_transformation}, and derive the initial condition
\eqref{eqn:discrete_model_phi_ic_eq}. This suggests the following 2-node discretization of the above
flux boundary condition
\begin{align}
&[(b-\theta^m_0) + (b-\theta^m_1)](R^m -\beta) = 
\nonumber \\
&2 ab + \frac ab \big[ (b-\theta^m_0)^2 + (b-\theta^m_1)^2\big] - 
\frac {4a} {\epsilon} \frac {\theta^m_1-\theta^m_0}{(b-\theta^m_0) + (b-\theta^m_1)},
\label{eqn:2node_CN_fluxBC}
\end{align}
however this is only second-order accurate at $n=1/2$. The 3-node boundary condition
\begin{align}
&(b-\theta^m_0)(R^m -\beta) = 
 ab + \frac ab (b-\theta^m_0)^2 - 
\frac a {2 \epsilon} \frac {-\theta^m_2+4 \theta^m_1 - 3 \theta^m_0}{(b-\theta^m_0)},
\label{eqn:3node_CN_fluxBC}
\end{align}
is second-order accurate at $n=0$ and generally appears to perform better than
\eqref{eqn:2node_CN_fluxBC}. We will exclusively use \eqref{eqn:3node_CN_fluxBC} for our flux
boundary condition in the computations to follow. The initial value of $\theta^0_0$ that matches
$R(0) = R^0$ is determined from \eqref{eqn:3node_CN_fluxBC} with $\theta^0_1 = \theta^0_2 = 0$.

Starting with known values $\theta^m_n$ for $n=0, 1, \ldots , N$; equations \eqref{eqn:NL_CN} for $n
= 1,2, \ldots ,N-1$, equation \eqref{eqn:3node_CN_fluxBC}, and $\theta^{m+1}_N = 0$, constitute a
coupled nonlinear system of equations for the determination of $\theta^{m+1}_n$ ($n=0, 1, \ldots ,
N$). A natural approach to the solution of these equations is to linearize and iterate. As such we
can write equations \eqref{eqn:NL_CN} and \eqref{eqn:3node_CN_fluxBC} in the form

\begin{align}
&2 \frac {\theta^{m+1}_{n,j+1} - \theta^m_{n}}{\delta} = \left[ \frac ab - \frac{ ab}{(b-\theta^m_n)^2}\right] \frac{\theta^{m}_{n+1} - \theta^{m}_{n-1}}{2\epsilon} 
\nonumber \\
&  \qquad + 
\frac a {\epsilon^2} 
\frac{\theta^{m}_{n+1} - \theta^{m}_{n}}{(b-\theta^m_{n+1})(b-\theta^m_n)} -\frac{\theta^{m}_{n} - \theta^{m}_{n-1}}{(b-\theta^m_{n})(b-\theta^m_{n-1})}
\nonumber \\
& \qquad +
\left[ \frac ab - \frac{ ab}{(b-\theta^{m+1}_{n,j})^2}\right] \frac{\theta^{m+1}_{n+1,j+1} - \theta^{m+1}_{n-1,j+1}}{2\epsilon} 
\nonumber \\
& \qquad  + 
\frac a {\epsilon^2} 
\frac{\theta^{m+1}_{n+1,j+1} - \theta^{m+1}_{n,j+1}}{(b-\theta^{m+1}_{n+1,j})(b-\theta^{m+1}_{n,j})} -\frac{\theta^{m+1}_{n,j+1} - \theta^{m+1}_{n-1,j+1}}{(b-\theta^{m+1}_{n,j})(b-\theta^{m+1}_{n-1,j})},
\label{eqn:NL_CN_it}
\\
&(b-\theta^{m+1}_{0,j+1})(R^{m+1} -\beta) = ab + \frac ab (b-\theta^{m+1}_{0,j+1})(b-\theta^{m+1}_{0,j})
\nonumber \\
&\qquad  - 
\frac a {2 \epsilon} \frac {-\theta^{m+1}_{2,j+1}+4 \theta^{m+1}_{1,j+1} - 3 \theta^{m+1}_{0,j+1}}{(b-\theta^{m+1}_{0,j})}.
\label{eqn:3node_CN_fluxBC_it}
\end{align}
Here instances of $\theta^{m+1}_n$ have been replaced by the approximations $\theta^{m+1}_{n,j}$
where $j = 0, 1, 2,\ldots$. The initial value of these approximations can be taken to be equal to
the corresponding values at the previous time step: $\theta^{m+1}_{n,0} = \theta^{m}_{n}$. Now given
known $\theta^{m}_{n}$ and $\theta^{m+1}_{n,j}$ values, quantities $\theta^{m+1}_{n,j+1}$ can be
calculated by solving a linear system almost identical in structure to \eqref{eq:linsys}. In
practice we observe that the $\theta^{m+1}_{n,j}$ values converge quite quickly, and we have adopted
a termination criteria
\begin{equation}
\max_{n} \left| \theta^{m+1}_{n,j+1} - \theta^{m+1}_{n,j}  \right| < 10^{-10},
\end{equation}
after which the final $\theta^{m+1}_{n,j+1}$ values are accepted and the calculation proceeds to the next timestep.

It is now clear that the above process of iteration implies a greater computational burden than
using the CN scheme for our integrable model. As the integrable model involves solution of a linear
differential equation, only one linear system of equations \eqref{eq:linsys} needs to be solved to
obtain exact $\phi^{m+1}_n$ values from known $\phi^{m}_n$ values. Conversely, when using iteration
to implement the CN scheme on equation \eqref{eqn:continuous_model_theta}, several linear systems of
comparable difficulty to \eqref{eq:linsys} may need to be solved to obtain sufficiently accurate
values of $\theta^{m+1}_{n,j}$ beginning from known $\theta^{m}_n$.

As such, we can also consider direct use of the CN scheme on the equation for $\theta$ without
iteration of the linearized equations \eqref{eqn:NL_CN_it}, \eqref{eqn:3node_CN_fluxBC_it} at each
timestep. That is, we simply accept the $\theta^{m+1}_{n,1}$ values as the final values of
$\theta^{m+1}_{n}$. By eliminating iteration, the CN scheme applied directly to the equation
\eqref{eqn:continuous_model_theta} produces a solution at time $M \delta $ after the solution of
only $M$ linear systems of equations --- just as the integrable CN model does.

\begin{figure}[ht]
\begin{tabular}{ll}
\includegraphics[scale=\mypicscale]{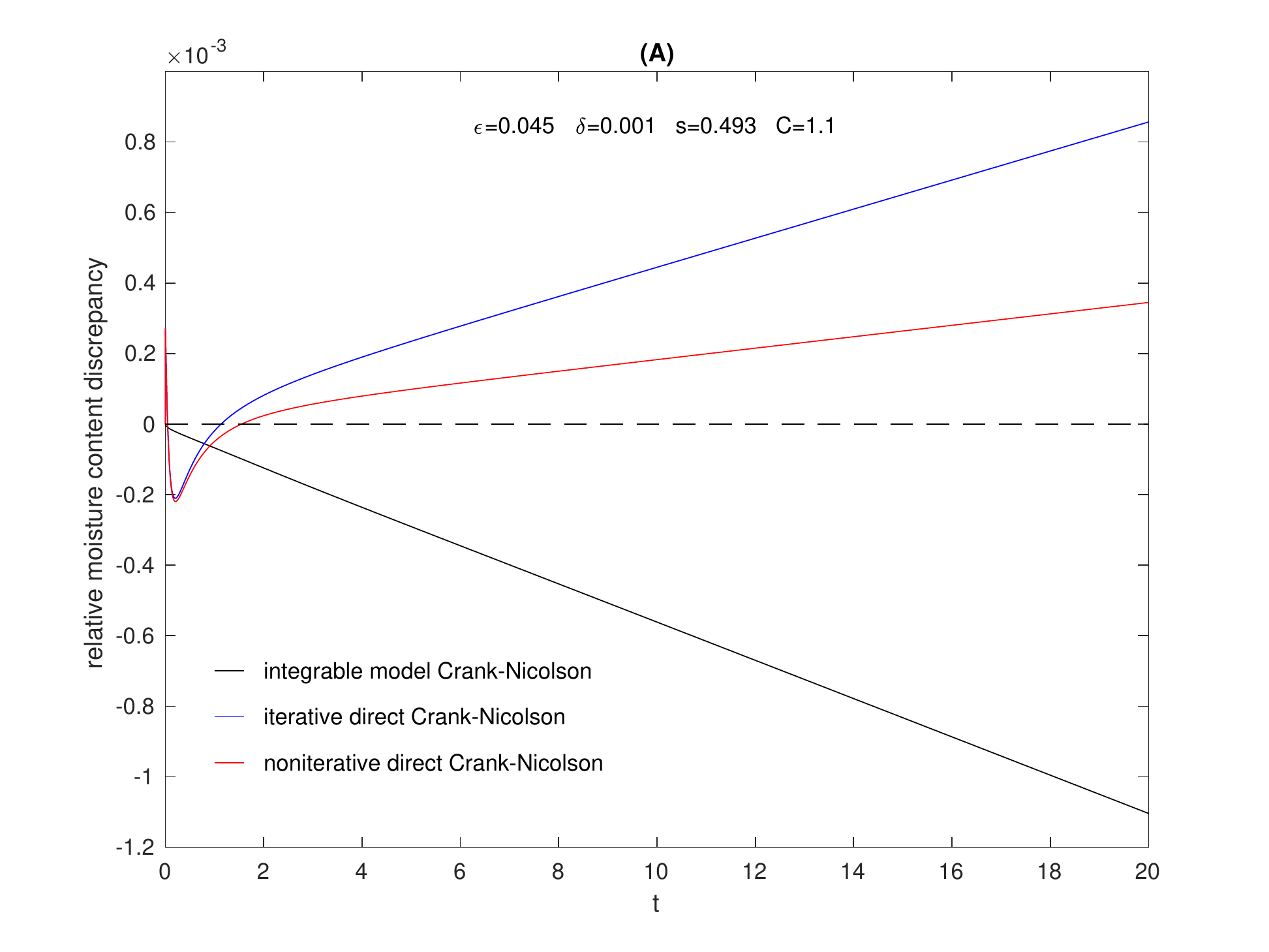}  & \hspace{\mypichin}
\includegraphics[scale=\mypicscale]{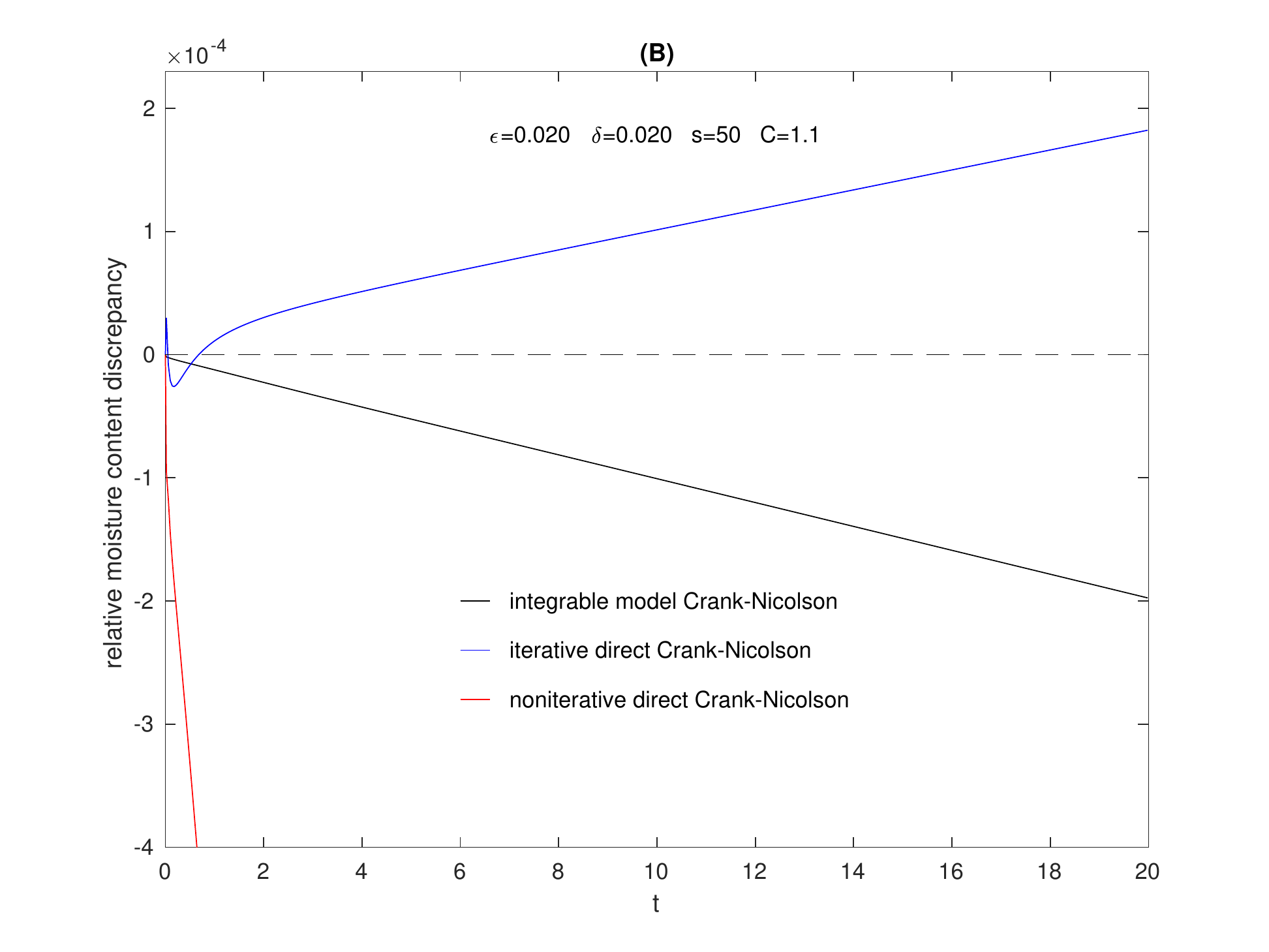} \vspace{\mypicvin}\\
\includegraphics[scale=\mypicscale]{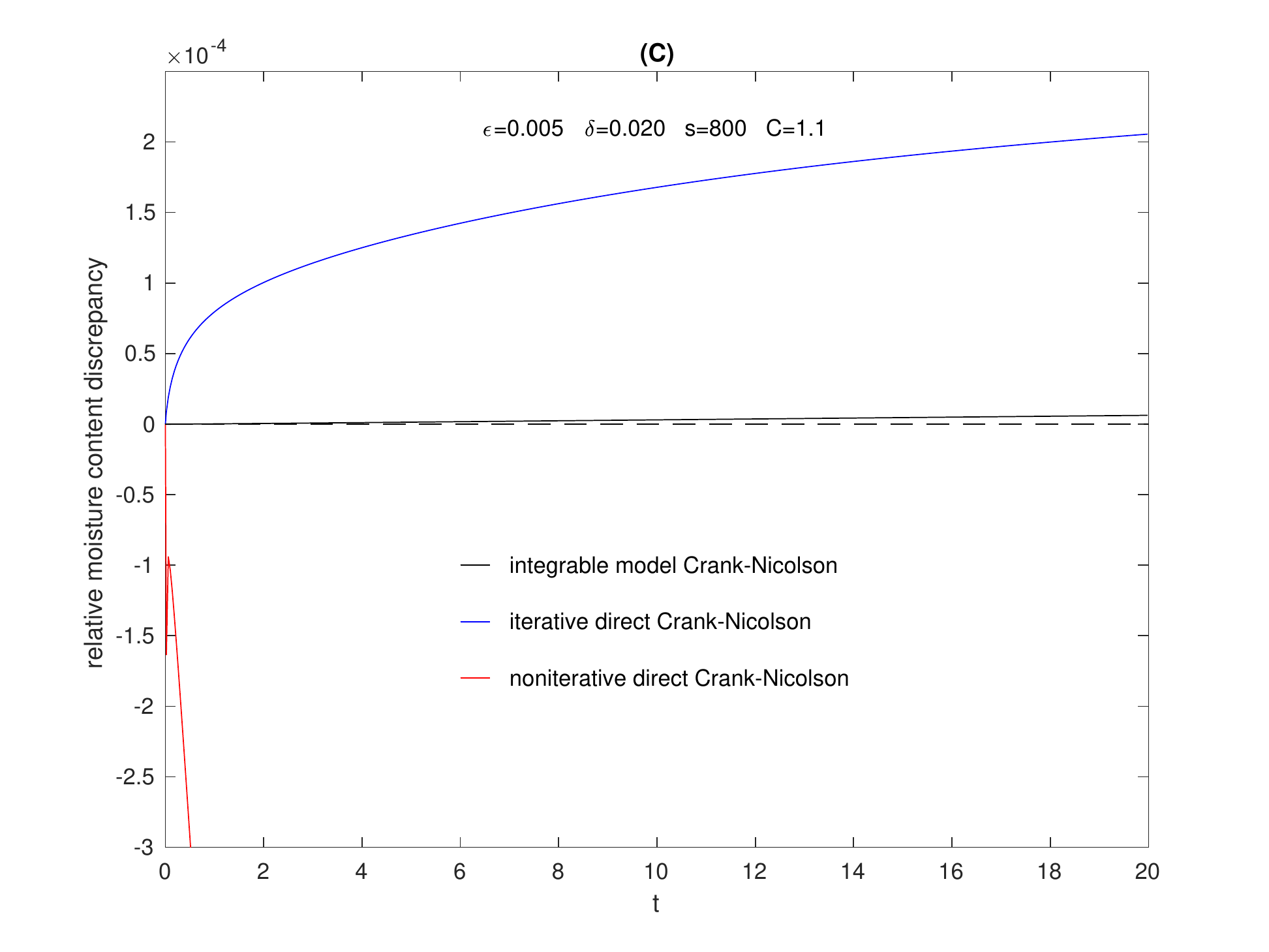} & \hspace{\mypichin}
\includegraphics[scale=\mypicscale]{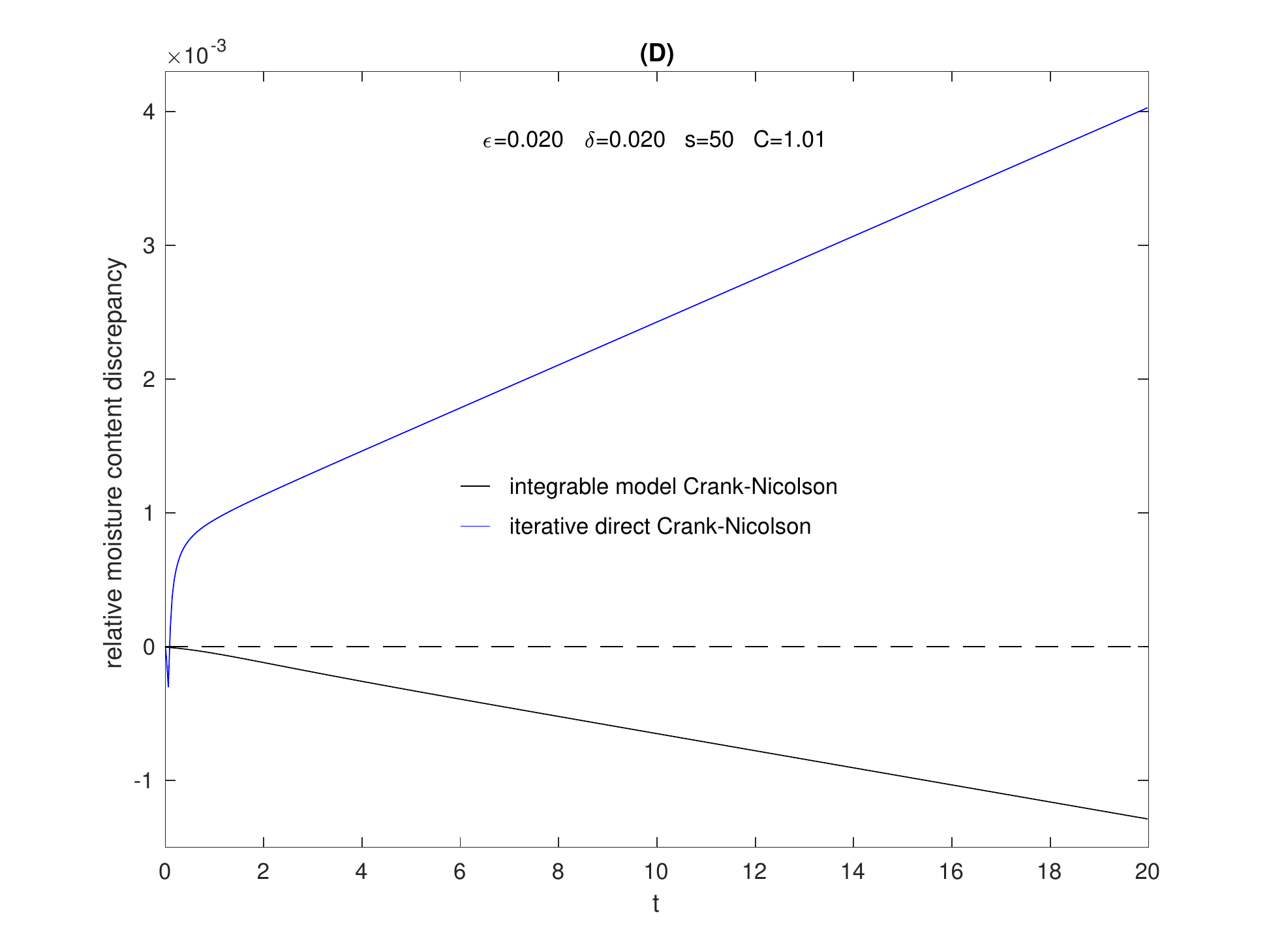}
\end{tabular}
\caption{Comparison of conservation of mass: The CN scheme applied directly to equation
\eqref{eqn:continuous_model_theta} for $\theta$ with or without iteration, compared to the
integrable CN model. In all cases $R(t) =0.6$ and $\theta(z,0) = 0$.  } \label{fig:Comparison}
\end{figure}

In addition to considering the limiting value of $\theta^m_0$, we can also evaluate the accuracy of
our numerical methods by comparing conservation of mass. With initial condition $\theta(z,0) =0$,
the relative moisture content discrepancy:
\begin{equation}
\frac{\displaystyle \frac 12 \sum_{n=0}^{N-1} (\theta^m_n + \theta^m_{n+1})(z^{m}_{n+1}- z^{m}_{n}) - \int_0^{m\delta } R(t) dt}
{\displaystyle \int_0^{M\delta } R(t) dt}
\label{eqn:moisture_discrepancy}
\end{equation}
should be of small magnitude at all times for accurate numerical schemes. Figure \ref{fig:Comparison} shows the relative moisture content discrepancy for $R =0.6$. In all cases $t_{\rm max} = \tau_{\rm max} = 20 = M \delta$; $Z_{\rm max} = 15 = N \epsilon$ for the integrable CN model and $z_{\rm max} = 15 = N \epsilon$ for the direct CN methods. In panels (B) and (C) the noniterative direct CN scheme approaches a final relative moisture content discrepancy $\simeq -10^{-2}$. In panel (D) the noniterative direct CN methods fails altogether. 

In panel (A) of Figure \ref{fig:Comparison} the iterative direct CN scheme involved the solution of
about 80000 linear systems similar to \eqref{eq:linsys}, compared to 20000 for the noniterative
direct and integrable CN schemes. In panels (B) and (C) the iterative direct CN scheme involved
solution of approximately 8000 linear systems compared to 1000 for the noniterative direct and
integrable CN schemes. Table \ref{tab:thetalim} shows the computed values of $\theta(0,20)$ for the
simulations of panels (A)--(C) of Figure \ref{fig:Comparison}.
\begin{table}[htp]
\begin{center}
\caption{$\theta(0,20)$ values for the simulations of panels (A)--(C) of Figure \ref{fig:Comparison}. \newline As $t \to \infty$, it is known that $\theta(0,t) \to 0.9496836.$ }
\vspace{0.1cm}
\begin{tabular}{|l||c|c|c|}
\hline
Crank--Nicolson  &$\epsilon = 0.045$  &$\epsilon = 0.020$  & $\epsilon = 0.005$\\
implementation	&$\delta = 0.001$	&$\delta = 0.020$	&$\delta = 0.020$\\
\hline 
integrable & 0.9496914  &0.9496845  &0.9495624 \\
\hline
iterative direct &  0.9496829 & 0.9496828 & 0.9496254 \\
\hline
noniterative direct  &  0.9496828  &  0.9496824 & 0.9488505\\
\hline
\end{tabular}
\end{center}
\label{tab:thetalim}
\end{table}

Equation \eqref{eqn:continuous_model_theta} aproaches a singular nonlinear limit as $C \to 1$, and
the value $C=1.01$ in panel (D) of Figure \ref{fig:Comparison} could physically represent a coarse,
sandy soil. As our governing equation becomes more nonlinear, the benefits of the integrable method
increase. The iterative direct CN method in panel (D) involved solution of approximately 21000
linear systems, compared to only 1000 for the integrable CN method. With $C=1.01$ and $R(t) = 0.6$;
$\theta(0,t) \to 0.9935477$ as $t \to \infty$. Computation using the iterative direct CN method
resulted in $\theta(0,20) = 0.9935478$, while the integrable CN method produced $\theta(0,20) =
0.9935476$.

Figure \ref{fig:Comparison} shows that direct applications of the CN scheme tend to exhibit sudden
changes in the moisture content discrepancy for early times, while the integrable CN model does
not. This is consistent with the self-adaptive moving mesh of the integrable model providing greater
accuracy for small values of $\theta$, at early times when the moisture content is increasing most
rapidly.

Overall, the integrable model clearly outperforms the noniterative direct CN method, and its
performance is broadly comparable to the iterative direct CN method.  As detailed above, the
iterative direct CN method involves a significantly greater computational effort, that yields no
clear benefit when compared to the results of the integrable CN model.

\section{Conclusion}\label{sec:Concluding_Remarks}

In this paper we considered an integrable model of one-dimensional groundwater infiltration, a
special case of the Richards equation. It takes the form of a nonlinear convection-diffusion
equation with time-dependent flux boundary conditions. For the special soil model considered,
the Richards equation can be transformed to the Burgers equation and the linear heat equation with an
additional convective term incorporating the known surface flux.

We have constructed integrable discrete models preserving the linearizability structure above, the
crucial components of this are discretization of the linear equation, as well as discretization of
the Cole-Hopf and Storm transformations. Three models have been presented. The first is based on the
naive Euler scheme often used in the theory of discrete integrable systems
\cite{HerederoLeviWinternitz1999,Hirota:difference5,Nishinari_Takahashi:Burgers}, which suffers from
built-in numerical instability based on the value of $s=\delta/\epsilon^2$. This is not suitable for
accurate computations of volumetric soil-water content. The second model is based on the discrete
Burgers equation which is a nonlinearization of the Euler scheme of the first model. This inherits
the numerical instability despite nonlinearization and again cannot be used for accurate
calculations. Finally we propose a model based on a stable, second-order Crank--Nicolson
discretization of our linear convection-diffusion equation.

We have assessed the performance of this integrable Crank--Nicolson model by comparing against the
Crank--Nicolson method directly applied to the original nonlinear convection--diffusion
equation. The accuracy of the final solution computed using a variety of lattice parameters was
observed to be approximately equal. However, directly applying the Crank--Nicolson method to the
nonlinear equation was found to involve significantly more computational effort than the integrable
model as measured by the number of linear systems solved during numerical integration.

To further improve computational performance, higher-order numerical integration schemes such as
higher-order Runge--Kutta methods could be applied to the transformed linear convection diffusion
equation, while suffering some inconvenience in the form of more elaborate boundary condition
implementation. As demonstrated for the Crank--Nicolson scheme, we expect that such higher-order
integrable models will exhibit similar reductions in computational cost compared to applying the
relevant higher-order schemes directly on the original nonlinear equation. As observed in this
study, the reward for exploiting the integrable nature of the original equation should increase as
the parameter $C$ decreases, and the nonlinearity of the original equation becomes more severe.

\section*{Acknowledgments}
The work has been done as an activity of the Kyushu University, Institute of Mathematics for
Industry (IMI), Australia Branch, which is managed with generous support from Kyushu University, La
Trobe University and the Ministry of Education, Culture, Sports, Science and Technology (MEXT),
Japan.  This work has been partially supported by JSPS KAKENHI Grant Numbers JP15K04909, JP16H03941,
JP16K13763. \mbox{K.~Maruno} was also supported by JST CREST. \mbox{P.~Broadbridge} gratefully acknowledges
support from a La Trobe Asia Visiting Fellow Grant, and JSPS Invitation Fellowship in Japan
(Short-term) S15706. 

\end{document}